\documentclass[a4paper,twoside]{article}
\newcommand{\affil}[1]{$^{\rm #1}$}
\usepackage{txfonts}
\usepackage{lmodern}
\usepackage{graphicx}
\usepackage{placeins}
\usepackage{subfig}
\date{} %Please leave the date blank
\title{\large\bf\flushleft Tidal stability of giants molecular clouds in the Large Magellanic Cloud}
\author{\parbox{\textwidth}{\flushleft
\vspace{-0.5cm}
{\it E. Thilliez\affil{1,2,5}, S.T. Maddison\affil{1}, A. Hughes\affil{3}, T. Wong\affil{4} }\\
\vspace{0.4cm}
{\small \affil{1}\,Centre for Astrophysics \& Supercomputing, Swinburne University of Technology, Hawthorn, VIC 3122, Australia}\\
{\small \affil{2}\,Universit\'{e} Paris-Sud 11, 91400 Orsay, France}\\
{\small \affil{3}\,Max-Planck-Institut f\"{u}r Astronomie, K\"{o}ningstuhl 17, D-69117, Heidelberg, Germany}\\
{\small \affil{4}\,Astronomy Department, University of Illinois, Urbana, IL 61801, USA}\\
{\small \affil{5}\,Email: ethilliez@swin.edu.au}}}
\begin{document}
\twocolumn[
\begin{changemargin}{.5cm}{.5cm}
\begin{minipage}{.9\textwidth}
\vspace{-1cm}
\maketitle
%
%
%%%%%%%%%%%%%     ABSTRACT    %%%%%%%%%%%%%
%Abstract of no more than 200 words here.
\small{\bf Abstract: Star formation does not occur until the onset of gravitational collapse inside giant molecular clouds. However, the conditions that initiate cloud collapse and regulate the star formation process remain poorly understood. Local processes such as turbulence and  magnetic fields can act to promote or prevent collapse. On larger scales, the galactic potential can also influence cloud stability and is traditionally assessed by the tidal and shear effects. \\
\bf In this paper, we examine the stability of giant molecular clouds (GMCs) in the Large Magellanic Cloud (LMC) against shear and the galactic tide using CO data from the Magellanic Mopra Assessment (MAGMA) and rotation curve data from the literature. We calculate the tidal acceleration experienced by individual GMCs and determine the minimum cloud mass required for tidal stability. We also calculate the shear parameter, which is a measure of a cloud's susceptibility to disruption via shearing forces in the galactic disk. We examine whether there are correlations between the properties and star forming activity of GMCs and their stability against shear and tidal disruption. \\
\bf We find that the GMCs are in approximate tidal balance in the LMC, and that shear is unlikely to affect their further evolution. GMCs with masses close to the minimal stable mass against tidal disruption are not unusual in terms of their mass, location, or CO brightness, but we note that GMCs with large velocity dispersion tend to be more sensitive to tidal instability. We also note that GMCs with smaller radii, which represent the majority of our sample, tend to more strongly resist tidal and shear disruption. Our results demonstrate that star formation in the LMC is not inhibited by to tidal or shear instability.}

%%%%%%%%%%%%%     KEYWORDS    %%%%%%%%%%%%%
\medskip{\bf Keywords:} Magellanic Clouds - ISM: clouds - instabilities - stars: formation
% Please write all keywords in lower case. PASA uses the
% standard list of subject headings adopted by The Astrophysical Journal
% and available from http://www.journals.uchicago.edu/ApJ/keywords_text.html.
% Keywords are separated by em-dashes, i.e. ---

%%%%%%%%DO NOT EDIT%%%%%%%%%%%%
\medskip
\medskip
\end{minipage}
\end{changemargin}
]
\small

\section{Introduction}
Star formation occurs in the densest regions of molecular clouds which undergo gravitational collapse. Therefore studying the evolution and stability of giant molecular clouds (GMCs) allows us to investigate which processes lead to star formation. There are many theories on how a GMC can collapse and form stars, but this topic remains poorly understood. To address this question we have to look at the forces acting on clouds, such as turbulence, magnetic fields and rotation.  Zuckerman $\&$ Evans (1974) and Fleck (1980) propose that turbulence inside GMCs is important and stabilises the cloud against gravitational collapse, and thus star formation is indirectly related to the turbulence~(Hennebelle $\&$ Chabrier 2011). However Ballesteros-Paredes et al. (2007) suggest that on smaller scales turbulence can also promote local collapse. Magnetic fields inside the cloud can delay collapse until ambipolar diffusion removes magnetic support~ (Shu 1985), and thus magnetic fields are also important in regulating star formation (Mouschovias et al. 2011).

The Large Magellanic Cloud (LMC) is one of the closest galaxies to the Milky Way, at a distance 50.1 kpc~(Alves 2004). The irregular shape of the LMC is commonly attributed to the tidal force exerted on it by the Milky Way~(Lin et al. 1995), and this interaction is assumed to be the main reason for the LMC's  episodic high star formation rate (Indu $\&$ Subramaniam 2011).Its metallicity is lower than the Milky Way (Westerlund 1997), and it thus represents a good analogy to high redshift galaxies. Its nearby location and near face-on orientation make it an ideal galaxy to study star formation, as we can accurately map and resolve the GMCs. Previous studies have focused on local effects such as turbulence or magnetic fields inside the interstellar medium in studying cloud stability and star formation. Since the LMC's irregular shape suggests that it is tidally stressed and deformed on a galactic scale, we propose to look for potential large-scale influences on star formation: the galactic tide and shear due to differential galactic rotation.

Given their mass and size, the GMCs of the Milky Way only exceed the minimum mass for tidal stability by a factor of 3, Stark $\&$ Blitz (1978) thus proposed that tidal forces can have an influence on GMC morphology. Similar conclusions were found by Blitz (1985) for the GMCs in M31. Rosolowsky $\&$ Blitz (2005) evaluated the tidal stability of GMCs in M64 and found them globally stable, except in the very inner part of the galaxy. Blitz $\&$ Glassgold (1982) found that whereas atomic clouds are in tidal balance in M101, the atomic clouds in the LMC are 5 times less massive than necessary to resist tidal disruption, and hence their resulting lifetime could be very short, of order $10^7$~years. Similarly, the study by~Ballesteros-Paredes et al. (2009) concluded that the Taurus molecular cloud in the Milky Way is also suffering significant galactic tidal disruption since its tidal energy is at least three times larger than its gravitational energy.

GMCs can also be disrupted by galactic rotational shear which can tear clouds apart. Large scale numerical simulations by Dobbs $\&$ Pringle (2013) suggest that clouds which are very filamentary can easily be torn apart. Using a multicomponent Toomre parameter to assess large scale effects on GMCs stability, Yang et al. (2007) recently concluded that gravitational instability of the disk drives most large-scale star formation in the LMC. However it has been demonstrated by Hunter et al. (1998) and Elson et al. (2012) that a shear parameter based on the time available for perturbations to grow inside the GMCs is a more efficient way to identify regions of nearby galactic disks that are actively star-forming.

The main purpose of this paper is to investigate if galactic tidal and shear effects in the LMC can influence GMC stability and star formation. We use the Magellanic Mopra Assessment dataset, which was a high angular resolution 12 CO (J=1-0) mapping survey of the Magellanic Clouds using the 22-m Mopra Telescope\footnote{The Mopra Telescope  is operated by the Australia Telescope National Facility (ATNF) which is a division of CSIRO.} This survey (Hughes et al. 2010; Wong et al. 2011) targeted initially the brightest clouds of the NANTEN survey (Fukui et al. 2008) to study the basic properties of giant molecular clouds. One of the main results of this survey is that most massive GMCs are associated with luminous young stars, but the limited sensitivity of the survey and lack of knowledge about stellar ages have prevented the determination of a characteristic GMC lifetime. Moreover, Wong et al. (2011) found no correlation between the virial parameter  -- the ratio of a cloud's kinetic to self-gravitational energy -- of LMC GMCs and the presence of young stellar objects, suggesting there may not be a connection between the global  stability of a GMC and its level of star-formation activity.
In this work, we start by defining the shear parameter and tidal stability, and then calculate their effects on the GMCs for a range of possible LMC rotation curves. Then we investigate the stability state of the GMCs and search for correlations between cloud properties such as velocity dispersion, morphology, position in the galaxy and their stability state. Finally we discuss the potential link between shear, the galactic tide and star formation in the LMC.

%%%%%%%
\section{GMC stability in the galactic potential}
In this section, we describe the two methods that we use to assess the stability of GMCs. Both are large-scale dynamical effects describing the interaction between GMCs and the overall galactic potential.
\subsection{The effect of shear}
\subsubsection{Physical interpretation}
To characterize the impact of differential rotation on the stability of GMCs against gravitational collapse, the Toomre parameter is commonly used. It quantifies the competition between self-gravity, pressure and coriolis forces experienced by a cloud. However, to correctly trace the stability against shear and compare it with star formation activity, one must often derive a multicomponent Toomre parameter which includes the effects of the stellar potential (Yang et al. 2007) or even the dark matter potential (Elson et al. 2012), plus the gas potential.\\

Hunter et al. (1998) points out that the coriolis force assuming angular momentum conservation can be overestimated since angular momentum is expected to be carried away by the magnetic field. They instead proposes to use another quantity called the shear parameter: the idea is to evaluate the time available for perturbations to collapse in the presence of local rotational shear due to the global rotation of the galaxy. The shear parameter value can differ from the Toomre parameter from $-12~\%$ for a flat rotation curve to $-50~\%$ for a slow rising rotation curve. Elson et al. (2012) recently showed that using the shear parameter can better trace the star formation activity in the inner part of the dwarf galaxies NGC 2915 and NGC 1705 than a more complex multicomponent gas + star Toomre criterion.\\

\subsubsection{Shear parameter $S_{g}$}
Following the work of Dib et al. (2012) and Hunter et al. (1998), we derive the shear parameter. The timescale over which a density perturbation can grow effectively against galactic rotational shear is of order $1/A$, where $A$ is the Oort constant. With dimension $t^{-1}$, the Oort constant $A$ measures the local shear level and is defined by:
\begin{equation}
A=0.5 \left( \frac{V}{R} - \frac{dV}{dR} \right) \, ,
\label{eq:1}
\end{equation}
where $V$ is the galactic rotation velocity and $R$ the galactic distance. For a cloud of radius $r$, with a rotational velocity $V$ at a galactic distance $R$, $A$ becomes:
\begin{equation}
A= 0.5 \left( \frac{V}{R} - \frac{\mid V(R+\frac{r}{2}(1+g)) - V(R-\frac{r}{2}(1+g)) \mid}{2r} \right) \, ,
\label{eq:2}
\end{equation}
where $g$ is a gaussian distributed random number between $-0.5$ and $0.5$, with the mean value of zero. The addition of the $\pm gr/2$ terms represent the non-sphericity of the GMC, thereby modeling the extension of a filamentary cloud toward the inner center and outer part of the galaxy. This formula is similar to that derived in Dib et al. (2012). Thus, using a perturbation growth rate of $\pi G \Sigma / \sigma$, the amplitude of the growth from an initial density perturbation $\delta \Sigma_{0}$ against shear is given by:
\begin{equation}
\delta \Sigma_{peak} = \delta \Sigma_{0} \exp \left( \frac{2\pi G \Sigma}{\sigma A} \right) \, ,
\label{eq:3}
\end{equation}
where $\Sigma$ and $\sigma$ are the local gas surface density and velocity dispersion, and $G$ the gravitational constant. According to Dib et al. (2012), for a perturbation to be significant in terms of instability, it must grow by a factor of at least $C$ = $1000$. Thus $C$ represents the density contrast between of average cloud ($\sim 10^{2}$~cm$^{-3}$) and a strongly gravitationally bound cloud ($\sim 10^{5}$~cm$^{-3}$), which in our case is given by:
\begin{equation}
C=\frac{\delta \Sigma_{peak}}{\delta \Sigma_{0}}= \exp \left( \frac{2\pi G \Sigma_{sh}}{\sigma A} \right) \, .
\label{eq:4}
\end{equation}
With that in mind, we can easily derive the critical surface density, $\Sigma_{sh}$, which represents the minimal surface density needed to resist the shear, given by:
\begin{equation}
\Sigma_{sh}=\frac{A \sigma \ln(C)}{2\pi G} \, .
\label{eq:5}
\end{equation}
Finally we can define the shear parameter, $S_{g}$, which is the ratio between the critical surface density and the actual gas surface density:
\begin{equation}
S_{g}=\frac{\Sigma_{sh}}{\Sigma}=\frac{A \sigma \alpha_{A}}{\pi G \Sigma} \, ,
\label{eq:6}
\end{equation}
with $\alpha_{A} = \ln(C)/2 \sim$ 3.45. Shear will therefore be effective at tearing apart the cloud if $S_{g}>1$ and be ineffective if $S_{g}<1$. We will use this equation to evaluate the stability state of our GMC sample against galactic shear.

%%%

\subsection{The effect of tides}
\subsubsection{The tidal acceleration $T$}
To study the tidal stability of GMCs in the LMC, we begin by considering the local gravitational stability and then add the influence of the galaxy. To evaluate the stability state of the GMCs, we use the Roche criterion which defines the region where tidal forces dominate over gravity.

We initially consider clouds which are self-gravitating, or ``bound clouds" as defined by Blitz $\&$ Glassgold (1982). A cloud will be bound if its gravitational energy is greater than its kinetic energy, which is traced by the velocity dispersion, $\sigma$, via:
\begin{equation} 
|\frac{-GM}{r}| \geq \frac{1}{2}\sigma ^{2}  \, ,
\label{eq:7}
\end{equation}
where $M$ is the mass of the cloud and $r$ the radius of the cloud. When this criterion is met, internal turbulence will not disrupt an isolated cloud, which is said to be gravitationally stable.\\

Next we add the influence of the galaxy, which can contribute to the cloud stability in two ways: (i) it can act like gravity and help confine the cloud, and (ii) it can act against gravity and disrupt the cloud. The expression for the tidal acceleration $T$ given by Stark $\&$ Blitz (1978) is:
\begin{equation}
T=\frac{V^{2}}{R^{2}}-\frac{\partial}{\partial R}\left(\frac{V^{2}}{R}\right)  \, ,
\label{eq:8}
\end{equation}
which can be understood by considering two particles in a molecular cloud: one at the centre of the cloud at a distance $R$ from the galactic centre and the second at the edge of the cloud at a distance of $R+r$ from the galactic centre. As these two particles belong to the same molecular cloud, they orbit around the galactic centre with the same angular frequency (if not the cloud will become elongated during the orbit around the galactic centre). The acceleration of the first particle on the second required to maintain the cohesion of the cloud is $Tr$. 

If now we ignore the internal velocity of the cloud and integrate the cloud radius times the tidal acceleration $Tr$ along its radius, our bound cloud must obey the following energy equation in order to resist the galactic tide:
\begin{equation}
|\frac{-GM}{r}| \geq \frac{1}{2} Tr^{2} \qquad \textrm{with $\sigma$ = 0} \, .
\label{eq:9}
\end{equation}
We will use various rotation curves of the LMC from the literature to obtain $V(R)$ and hence determine the tidal acceleration $T$.
\begin{figure*}
\begin{center}
  \subfloat{\includegraphics[width=60mm,height=45mm]{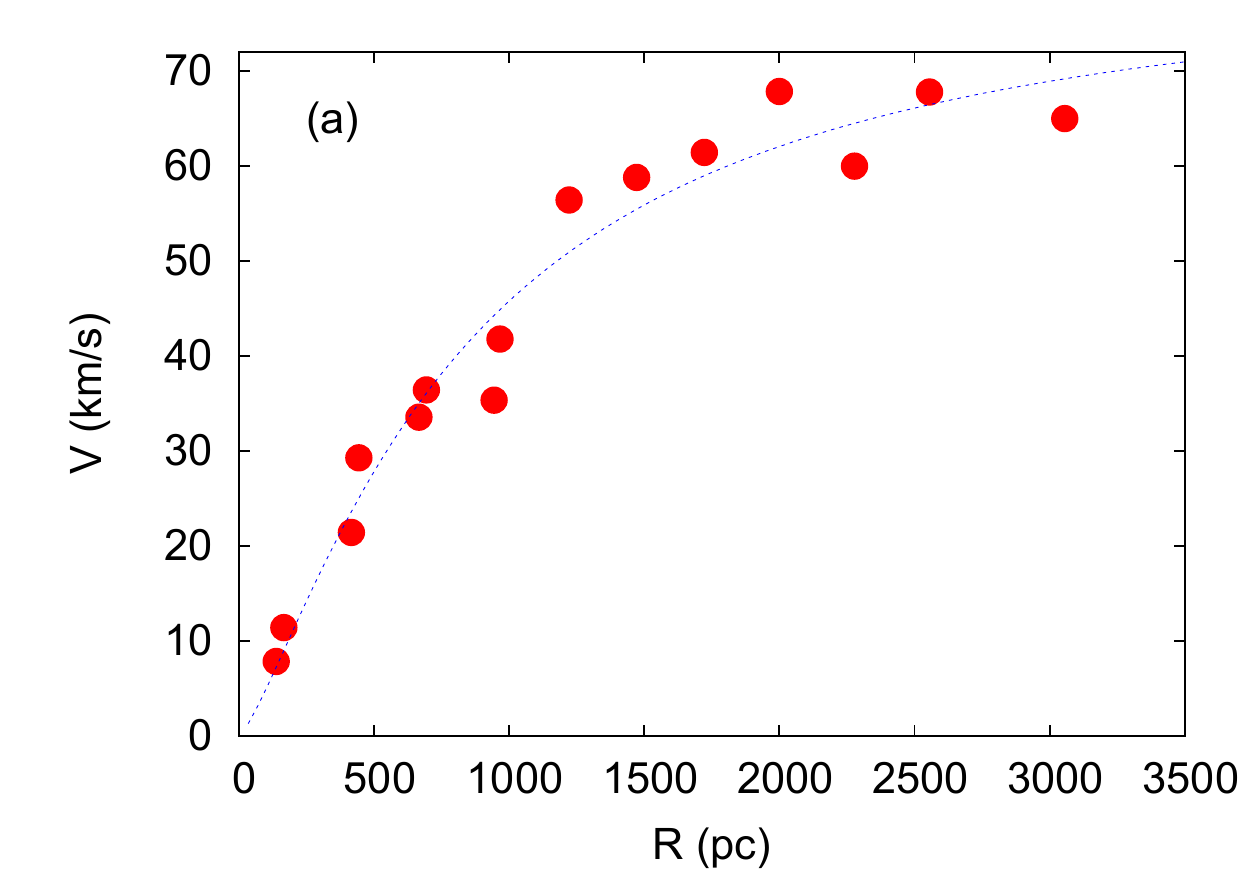}}                
  \subfloat{\includegraphics[width=60mm,height=45mm]{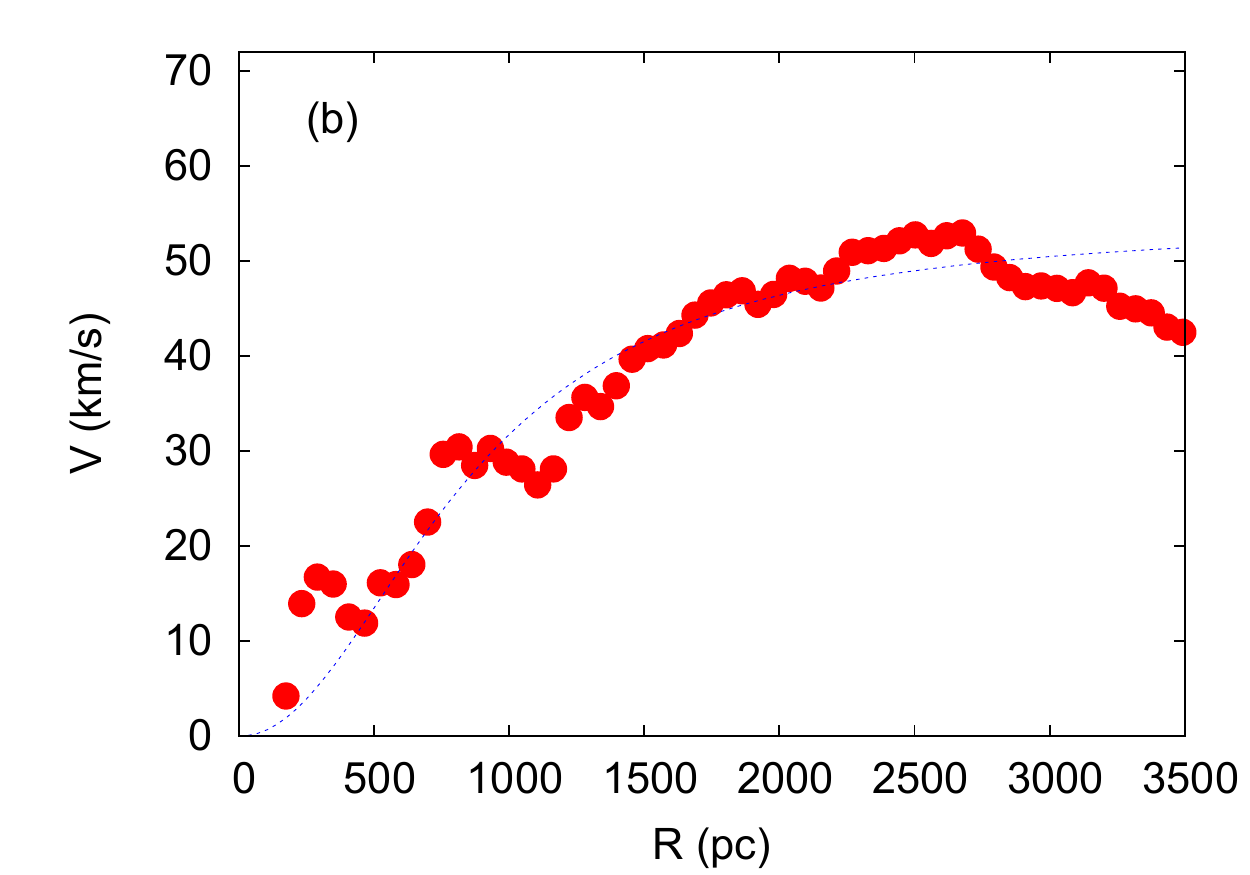}} 
  \subfloat{\includegraphics[width=60mm,height=45mm]{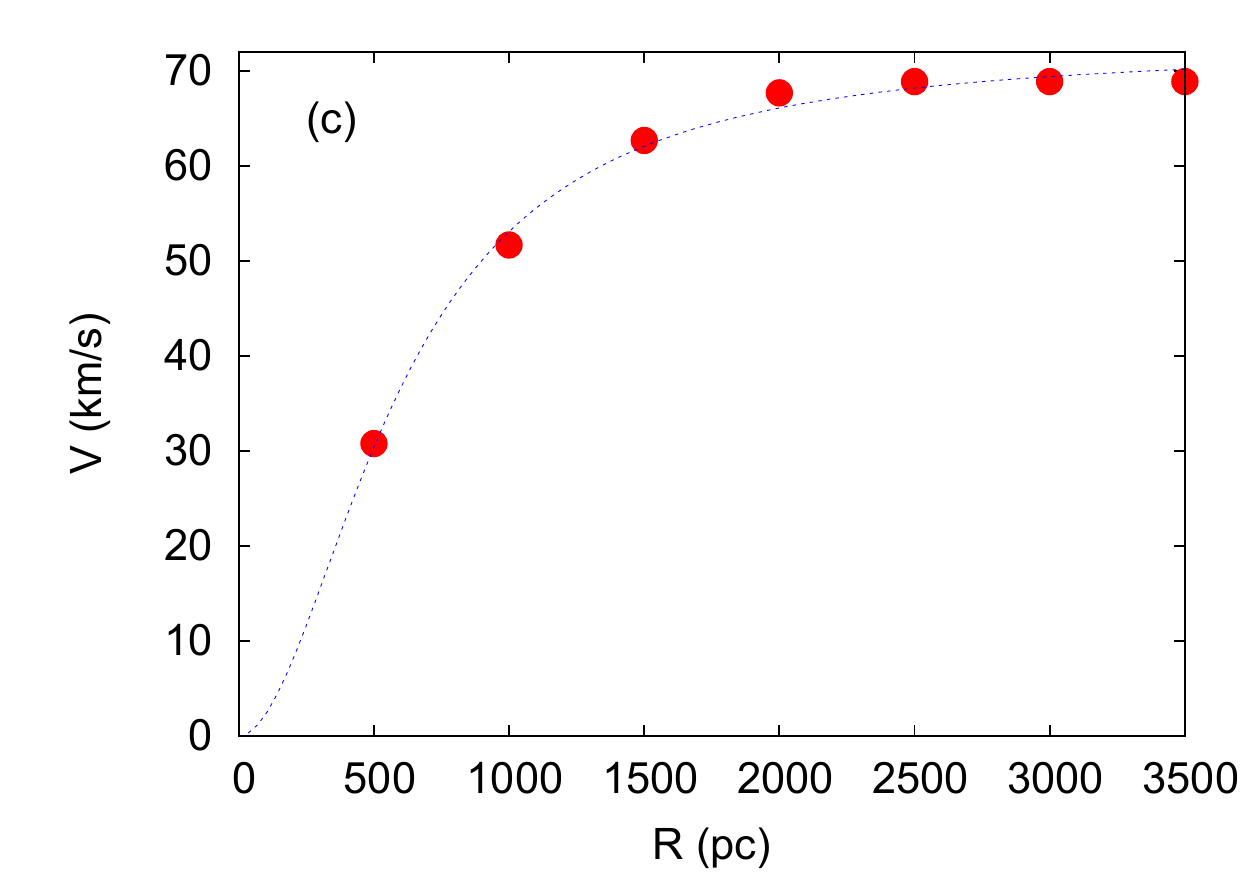}}                
  \caption{Least squares fit to the three LMC rotation curves, $V(R)$, used in this work from  (a) Feitzinger (1979),  (b) Wong et al. (2009), and (c) Alves $\&$ Nelson (2000). The symbols are the observations from those works and the curves are our best fit to the data.} 
\label{fig:1}  
\end{center}
\end{figure*}

%%%
\subsubsection{Roche criterion}
The Roche limit of a body orbiting around a more massive object is the minimal distance where this body is held together by its own gravity in the rotating frame. Below this limit, the orbiting body can be torn apart by the tidal force of the more massive object.
The Roche criteria for our GMCs is an equilibrium between the self gravity of the cloud, its internal pressure and the galactic tide.
To derive this expression, Stark $\&$ Blitz (1978) used the fact that the total energy (the sum of the kinetic, gravitational and tidal energy) of the cloud must always be less than the energy at the inner Lagrange point, a stationary position where the gravitational attraction of the galactic centre cancels the self-gravity of the cloud. If the total energy is higher than the energy at the inner Lagrange point, the cloud will migrate from its actual position to the inner Lagrange point and be torn apart around this point. Evaluating the potential at the inner Lagrange point, the Roche criterion is given by:
\begin{equation}
\frac{1}{2}\sigma ^{2} -\frac{GM}{r} - \frac{1}{2} Tr^{2} + \frac{3}{2}(GM)^{2/3}T^{1/3} < 0 \, .
\label{eq:10}
\end{equation} 
We will use this expression to determine the tidal stability of our sample of GMCs.

%%%%%%%%%%%%%%%%%%%%%%%%%%%%%%%%%%%%%%%%%%%%%%%%%%%%%%%%%%%%%%%%%%%%%%%%%%%%%%
\section{Data}
In this section we introduce the MAGMA survey and the catalog of the LMC GMCs used in this study and discuss how we evaluate the star formation activity in these clouds. We also describe the rotation curves used to estimate the galactic influence on the stability of GMCs.
\subsection{The MAGMA survey}
The Magellanic Mopra Assessment (MAGMA) consists of CO~(J=1-0) mapping survey of both the LMC and the Small Magellanic Cloud. Observations for the LMC were performed with the ATNF Mopra 22-m telescope from 2005 to 2010 and initially targeted the 114 brightest clouds of the NANTEN surveys by~ Fukui et al. (2008). The survey covered a total area of 3.6 deg$^{2}$, where the known CO complexes from the NANTEN survey were located, but with an improved angular resolution by a factor 4 to resolve the largest GMCs. The observed MAGMA fields cover a large fraction (80$\%$) of the total CO emission from the LMC traced  by the NANTEN data. The angular resolution of the MAGMA data is about $45^{''}$, corresponding to a linear scale of $\sim~11$pc. In this work, only clouds with a radius greater than $11$~pc were included in our analysis. The RMS of the noise fluctuations ranges from $0.2$ to $0.5$~K per $0.5$~km/s channel, with a mean value of $0.3$~K. For a typical GMC linewidth of $3$~km/s, the 3$\sigma$ sensitivity limit is 5~ M$_{\odot}/\rm pc^{2}$, assuming $X_{\rm CO}$=$2\times10^{20}$ cm$^{-2}$(K km s$^{-1}$)$^{-1}$.\\

To identify significant CO structures in the MAGMA data cubes, the automated CPROPS package (Rosolowsky $\&$ Leroy 2006) was used to identify ``emitting islands'', defined as isolated regions of significant emission (greater than 3$\sigma$) expanded in all directions in the data cube to a 2$\sigma$ edge. Those emitting regions were then decomposed into 450 individual clouds, of which 260 have $r>11$pc. The catalog derived by Wong et al. (2011) provides a range of information about the 260 resolved clouds in terms of  their structure and properties. For this work we use the GMCs' radius $r$, major-to-minor axes ratio $a/b$, galactocentric radius $R$, total integrated CO mass $M_{\rm CO}$, velocity dispersion $\sigma$, and peak CO brightness temperature $T_{\rm CO}$. The GMCs identified by MAGMA have typical radii between $5-71$~pc, masses between $3\times10^{3}-2\times10^{6}$ M$_{\odot}$ and velocity dispersion between $2-32$~ km/s, with average values of $28$~pc, $1.5\times10^{5}$ M$_{\odot}$, and $2.8$~km/s respectively. All the GMCs used in our study are located in the inner part of the LMC with $R < 3500$~pc. \\
%

%%%%%%%%%%%%%%%%%%%%%%%%%
\subsection{$24$ $\mu$m flux density}
It has recently been shown by~Relano $\&$ Kennicutt (2009) that there is a good spatial correlation between 24$\mu$m and H$\alpha$ emission in M33, suggesting that the dust emitting at 24~$\mu$m is predominantly heated by the emission coming from OB stars within the H II regions. Emission at 24~$\mu$m is thus an efficient signature of ionized photons inside an HII region. Therefore we will use the 24~$\mu$m surface brightness derived for each GMC of the MAGMA survey as a proxy for their star formation activity.\\

To trace the emission at 24~$\mu$m in the LMC, we use the {\it  Spitzer} mosaic obtained by the Surveying Agents of Galaxy Evolution (SAGE) Legacy Program (Meixner et al. 2006), using the Multiband Imaging Photometer (Rieke et al. 2004). The native angular resolution of this map is 6$^{''}$, and the surface brightness sensitivity to diffuse emission is 1 MJy sr$^{-1}$. We use the full enhanced LMC mosaic that is publicly available through the {\it  Spitzer} Science Centre archive\footnote{http://ssc.spitzer.caltech.edu/spitzermission/observingprograms/ legacy/sage/. The data delivery documentation, authored by M. Sewilo, is also  available at this URL.}. {\it Spitzer} observations for the SAGE project were scheduled at two different epochs, separated by an interval of three months, in order to minimize striping artefacts and to constrain source variability. We use the image that was produced by combining both epochs of observations. Processing of the SAGE Legacy data includes steps to remove residual instrumental signatures and to subtract background emission at 24~$\mu$m. To estimate the average 24~$\mu$m surface brightness associated with a GMC in the MAGMA catalogue, we use the mean value of all pixels within the projected area of cloud.\\

%%%%%%%%%%%%%%%%%%%%%%%%%%
\subsection{Galactic rotation curves}
In order to evaluate the tidal acceleration $T$ and the Oort constant $A$ for the shear parameter $S_{g}$, we need data from the rotation curve of the LMC which provides the radial velocity $V(R)$. Once the radial velocity is known, we can derive values for $T(R)$ and $A(R)$ throughout the LMC. Since both the tidal acceleration and the shear parameter are strongly dependent on the galactocentric radius $R$ (and by extension the radial velocity $V$) for small galactocentric radii, we evaluate the stability by using three different determinations of the LMC rotation curve. 
\begin{enumerate}
\item The first rotation curve is from Feitzinger (1979), which uses data from the HI survey of the LMC by~ McGee $\&$ Milton (1966). The beam size of the survey using was about $14.5^{'}$ and the velocity resolution $7$~km/s for a sample of 330 objects. This dataset was also used by Blitz $\&$ Glassgold (1982) to determine the tidal stability of atomic clouds in the LMC. The LMC parameters deduced by ~McGee $\&$ Milton (1966) and adopted by Feitzinger (1979) include distance (52 kpc), rotation center coordinates ($\alpha$= 5:20:00, $\delta$= -69:00:00), inclination ($27^{\circ}$) and position angle ($170^{\circ}$). 
\item The second rotation curve is from Wong et al. (2009), who use the HI survey of Kim et al. (1998) with improved proper motion measurements. The radial velocity map was corrected for the LMC's proper motion as determined by Kallivayalil et al. (2006) using the expressions given by van der Marel et al. (2002). The GIPSY program ROTCUR was used to fit a rotation curve with best fit values for the receding major axis position angle (341$^{\circ}$), the systemic velocity ($277$~km/s), and the center position ($\alpha$=5:19:30, $\delta$=-68:59:00) determined by Wong et al. (2009). The disk inclination, which is not well-constrained by the gas kinematics, was fixed at 35$^{\circ}$, based on the photometric study by van der Marel $\&$ Cioni (2001).
\item  The third rotation curve is that derived for carbon stars from Alves $\&$ Nelson (2000). Their procedure was as follows: positions and galactocentric radial velocities for Magellanic Cloud carbon stars were taken from Kunkel et al. (1997), discarding all Small Magellanic Cloud stars, intercloud carbon stars, and carbon stars located near the center of the LMC (Kunkel et al. 1997), resulting in homogeneous data set of 422 carbon stars. Then the carbon star velocities were corrected for the projected radial velocity gradient by adopting the LMC space motion calculated by Kroupa $\&$ Bastian (1997). Assuming the center of the LMC is at $\alpha$= 05:17:06 and $\delta$=-69:02:00 and an inclination of 33$^{\circ}$, they convert each carbon stars right ascension and declination into spherical coordinates, then derive the galactocentric radius $R$ and the velocity function $V$.
\end{enumerate}

A common parametrization of the radial velocity $V$ as a function of galactocentric radius $R$ is given by Binney $\&$ Tremaine (1987):
\begin{equation}
V=\frac{V_{0}}{1+(\frac{R}{R_{0}})^{-\gamma}} \, ,
\label{eq:11}
\end{equation}
where $V_{0}$, $R_{0}$ and $\gamma$ are free parameters. We used the method of least squares to determine these free parameters for the three rotation curves. We examined 340 values of $V_{0}$ ranging from $10-180$~km/s in even steps; 540 values of $R_{0}$ between $300-3000$~pc for each $V_{0}$; and a range of $\gamma$ values between $0.2-5$ in steps of 0.05 for each $R_{0}$ value. Thus more than $17\times10^{6}$ rotation curve models were generated, and each model was compared to the observational dataset. To find the best fit, we minimised the dispersion of the difference between the observed values and model values around an average. This was done by evaluating the $\chi^{2}$ value, which is the sum over all the $N$ data points of the difference between the observed velocity $V_{obs}(i)$, and modelled velocity $V_{mod}(i)$ divided by the observational uncertainty $\epsilon_{V}$ :
\begin{equation}
\chi^{2}= \sum_{i}^{N} \left( \frac{V_{obs}(i) - V_{mod}(i)}{\epsilon_{V}}\right)^{2} \, ,
\label{eq:12}
\end{equation}
where  $\epsilon_{V}$ is taken to be 7 km$/$s for fitting the curve of Feitzinger (1979), 2 km$/$s for ~Wong et al. (2009), and we used the errors provided by~Alves $\&$ Nelson (2000) in their Table 2 for $\epsilon_{V}$. We report the minimal values for $V_{0}$, $R_{0}$ and $\gamma$ in Table \ref{table:table1} and the corresponding rotation curves in  Figure \ref{fig:1}. We use these values to evaluate the tidal acceleration $T$ and the Oort constant $A$.
\begin{table}
\begin{center}
\caption{Rotation curve parameters from $\chi^{2}$ best fit to the three data sets.} 
\begin{tabular}{l c c c}\hline\hline
   Rotation curve & R$_{0}$ & V$_{0}$ & $\gamma$ \\
       & (pc) & (km/s) &      \\
  \hline
  Feitzinger (1979) & 835.0 & 82.0 &  1.30  \\
  Alves $\&$ Nelson (2000) & 592.0 & 72.5 & 1.92  \\
  Wong et al. (2009) & 845.0 & 54.0 & 2.10 \\
  \hline\hline
\label{table:table1}
\end{tabular}
\end{center}
\end{table}

%%%%%%%%%%%%%%%%%%%%%%%%%%%%%%%%%%%%%%%%%%%%%%%%%%%%%%%%%%%%%%%%%%%%%%%%%%%%%%%
\section{Results}
Here we report the results from our analysis using the shear parameter and the tidal acceleration to investigate the stability of the LMC's GMCs. In addition, potential relationships between these instability quantities and cloud properties, as well as star formation activity, are investigated.\\

We use the rotation curve from~Feitzinger (1979) to derive the values for the Oort constant values $A$ (equation \ref{eq:2}) and tidal acceleration $T$ (equation \ref{eq:8}). (Using the other two rotation curves does not strongly change our conclusions, but a qualitative discussion of their impact is presented in Section 5.) We assume a $X_{\rm CO}$ factor equal to $3\times10^{20}$ cm$^{-2}$ (K km s$^{-1}$)$^{-1}$ to derive the mass and surface density of the MAGMA GMCs from measurements of their CO luminosity and surface brightness, and then derive the ratio between the actual CO mass and the minimal mass, $M_{\rm CO}/M_{min}$, required for tidal stability. Ideally when calculating the GMC mass we would like to use the H$_{2}$ mass, since it is the major constituent of the GMCs. We discuss our choice of $X_{\rm CO}$ factor, and its potential impact on our results, in Section 5.\\  

\subsection{The shear parameter}
\begin{figure*}
\begin{center}
\includegraphics[width=100mm,height=75mm]{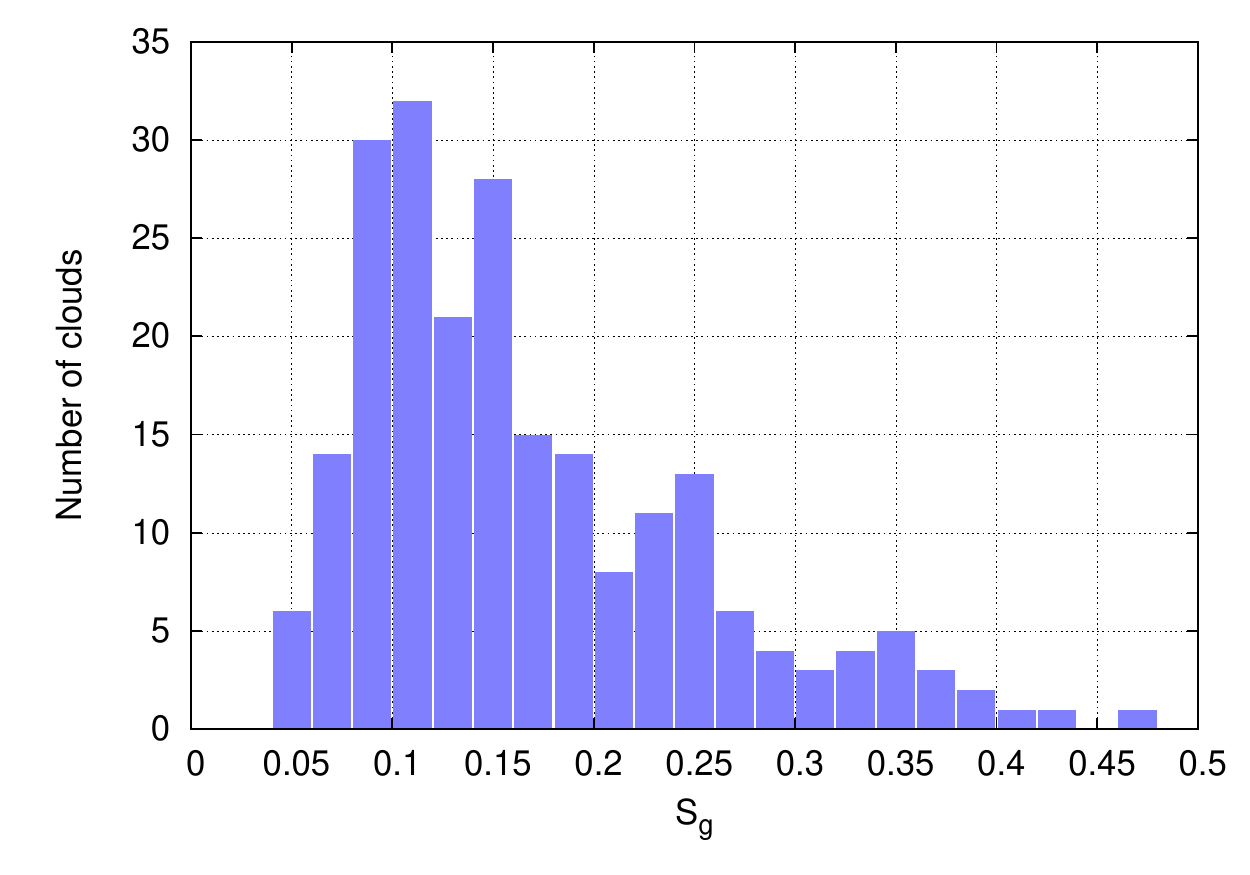}
\caption{Distribution of the shear parameter $S_{g}$ for our selected sample of 260 GMCs.}
\label{fig:2}
\end{center}
\end{figure*}

Figure \ref{fig:2} shows the distribution of the shear parameter, $S_{g}$, for all our GMCs. We can immediately see that the distribution ranges between 0.03 and 0.48 and peaks around $S_{g}\sim 0.12$. Since $S_{g}<1$, shear is not the dominant mechanism supporting the clouds against gravitational collapse. Similar conclusions were derived for the Milky Way by Dib et al. (2012). Note that we obtain an average value of the Oort constant $A=~13$ km/s/kpc, which is consistent with the local value for the Milky Way of $15$ km/s/kpc~(Feast $\&$ Whitelock 1997).
%Distribution of the Shear  parameter accross the galaxy
\begin{figure*}
\begin{center}
\includegraphics[width=100mm,height=75mm]{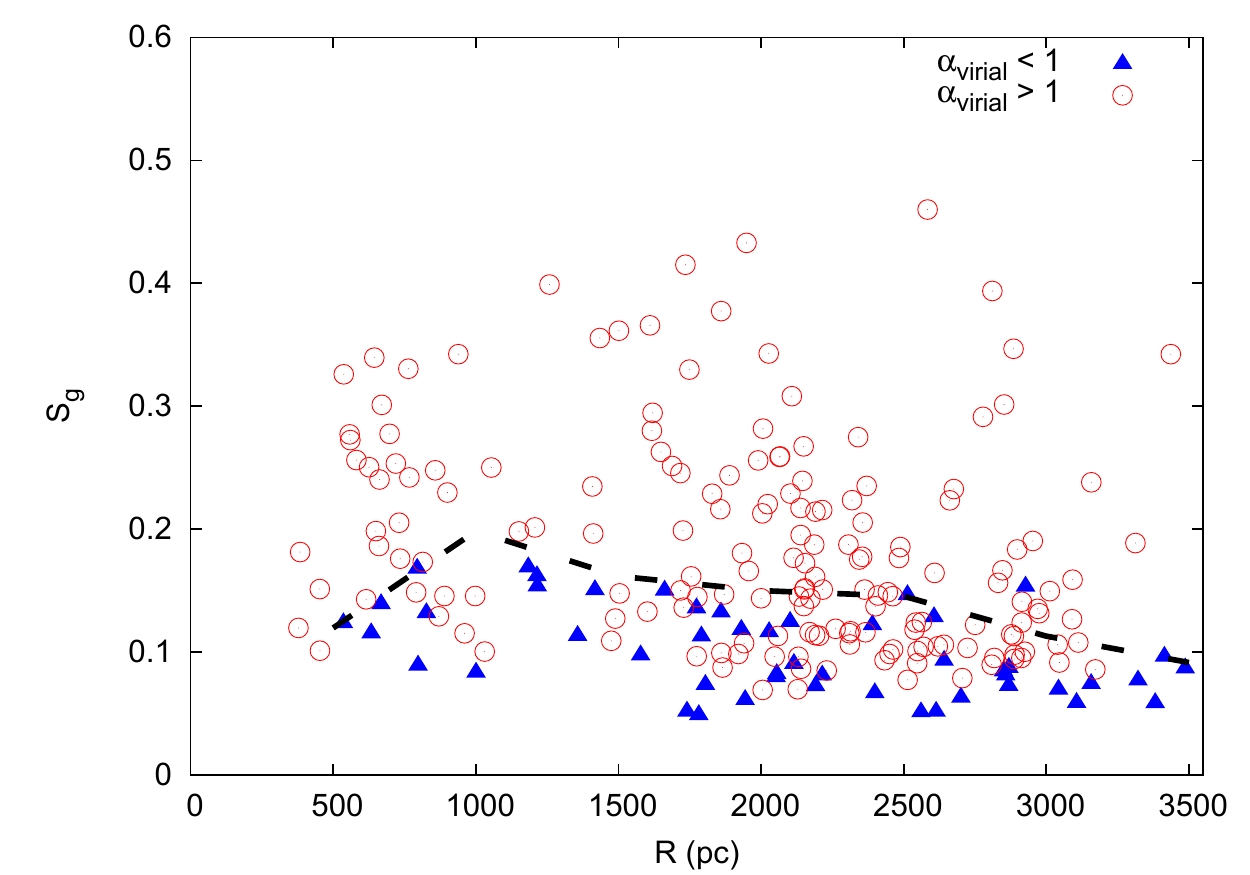}
\caption{Variation of the shear parameter, $S_{g}$, for our selected sample of 260 resolved GMCs as a function of the galactocentric radius, $R$. The blue triangles are the gravitationally bound clouds ($\alpha_{vir}$ $<$1) and the red dots are the unbound clouds ($\alpha_{vir}$ $>$1). The dotted line shows the median value of $S_{g}$ for the entire sample of clouds in radial bins of 0.5 kpc.}
\label{fig:3}
\end{center}
\end{figure*}
Moreover, as can be seen in Figure~\ref{fig:3}, the shear parameter $S_{g}$ remains almost constant (with a median value $\sim$~0.12) across the galaxy for gravitationally bound and unbound GMCs as defined by the virial parameter $\alpha_{vir}$, and thus no specific location with high values of the shear parameter $S_{g}$ could be identified. This result further supports the findings of Dib et al. (2012) for the Milky Way. We also investigated the importance of shear in the very high star forming region 30 Doradus (Torres-Flores et al. 2013), by selecting all the clouds in a 0.5$^{\circ}$ radius around the center of 30 Dor (RA: 84.658$^{\circ}$ and Dec: -69.095$^{\circ}$). We found eight clouds inside this radius with shear parameter, $S_{g}$, ranging between $0.09-0.32$ with a  mean value of $0.18$, which is consistent with the mean value over the complete sample $S_{g} \sim~0.12$. The virial parameter taken here as $\alpha_{vir}$=$5 r \sigma^{2}/ G M_{\rm CO}$ distinguishes the unbound clouds for which the kinetic energy dominates over the gravitational energy ($\alpha_{vir}>1$), from the gravitationally bound clouds which have less kinetic energy than gravitational energy ($\alpha_{vir}<1$). We notice that the sub-sample for unbound clouds ($\alpha_{vir}>1$) in Figure~\ref{fig:3} tend to have higher $S_{g}$ than the gravitationally bound clouds ($\alpha_{vir}<1$). While it is physically intuitive that less strongly bound clouds are more liable to disruption by external effects, we note that this correlation may be partially algebraically imposed, since $\sigma$ and $r$ both appear in the numerator of our expressions for $\alpha_{vir}$ and $S_{g}$. \\

%None correlation plots between Shear and SF, Peak CO brightness, shapes etc...
\begin{figure*}
\begin{center}
\subfloat{\includegraphics[width=90mm,height=67mm]{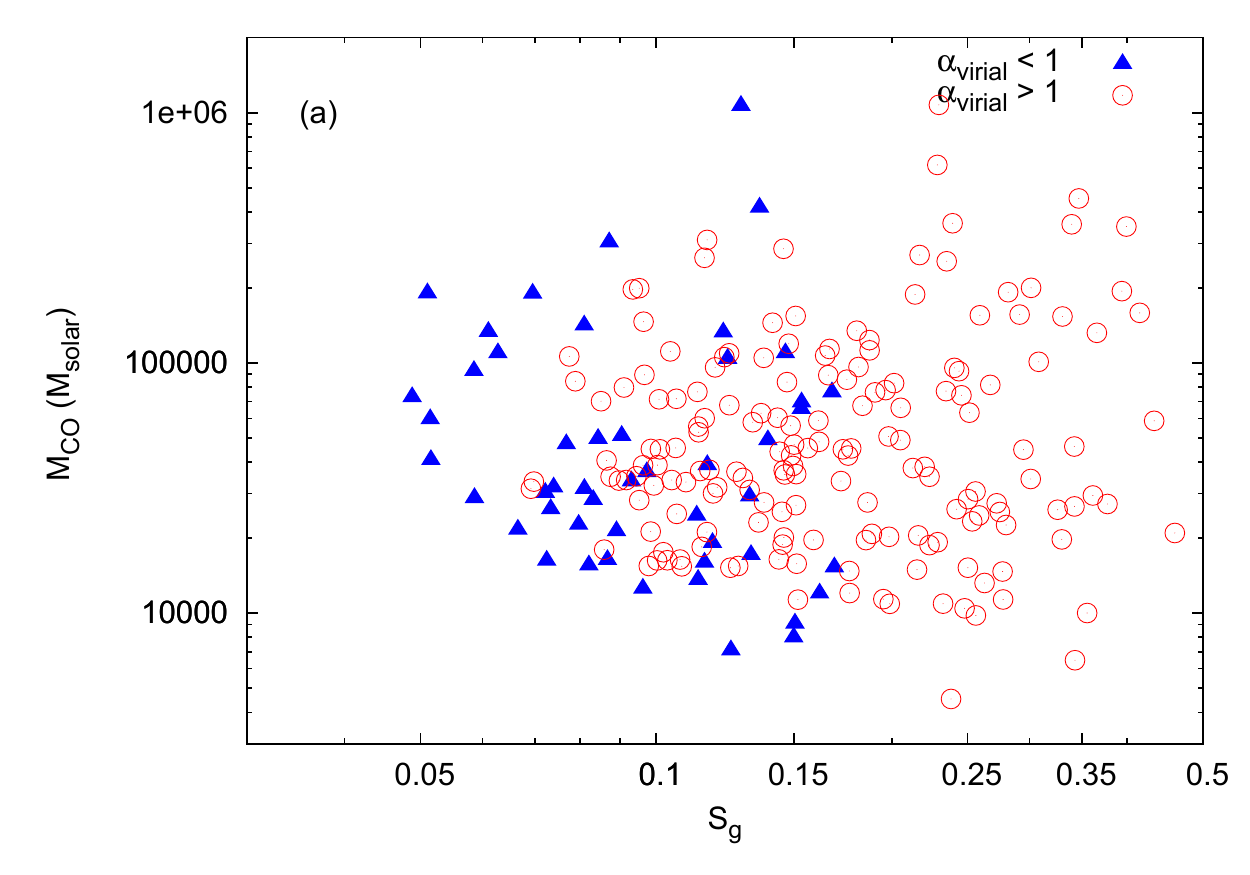}}                
\subfloat{\includegraphics[width=90mm,height=67mm]{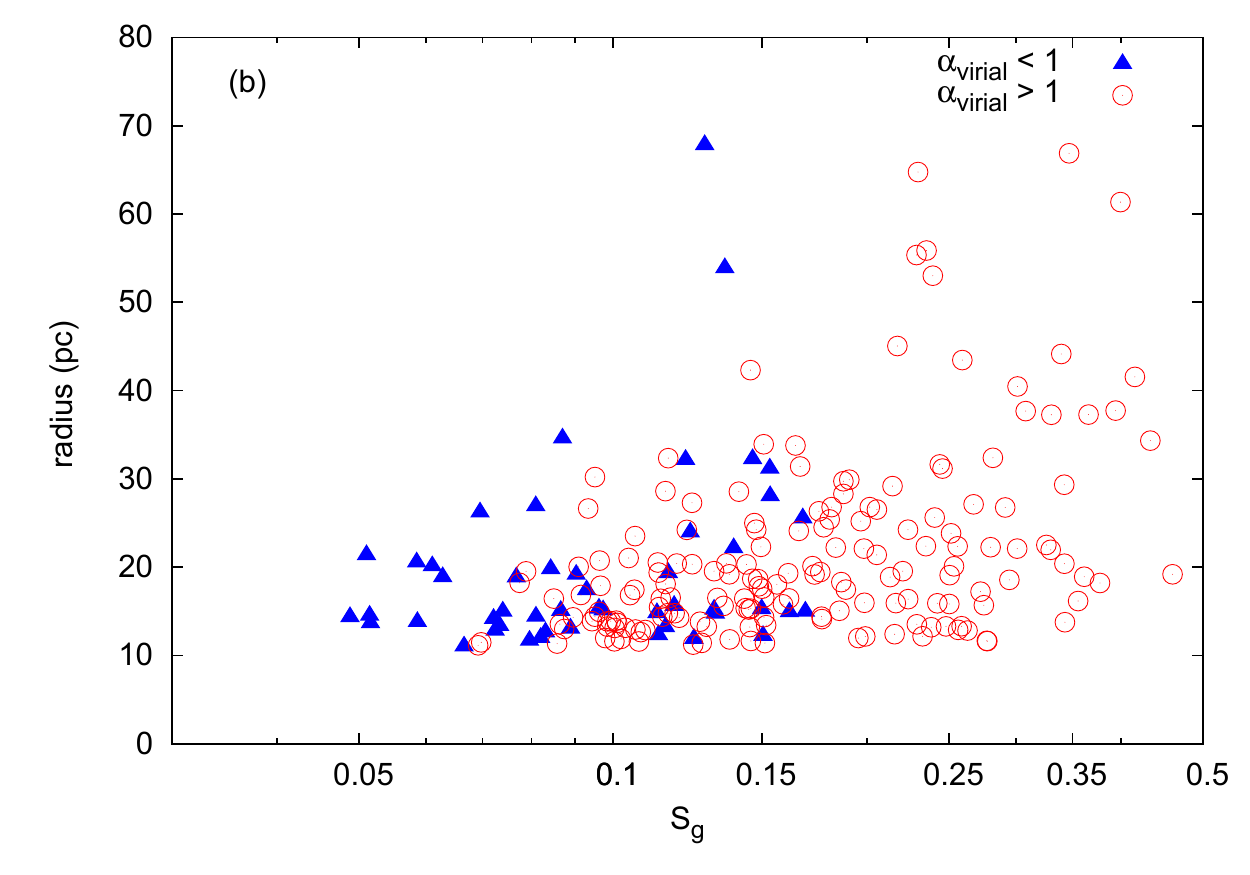}}\\     
\subfloat{\includegraphics[width=90mm,height=67mm]{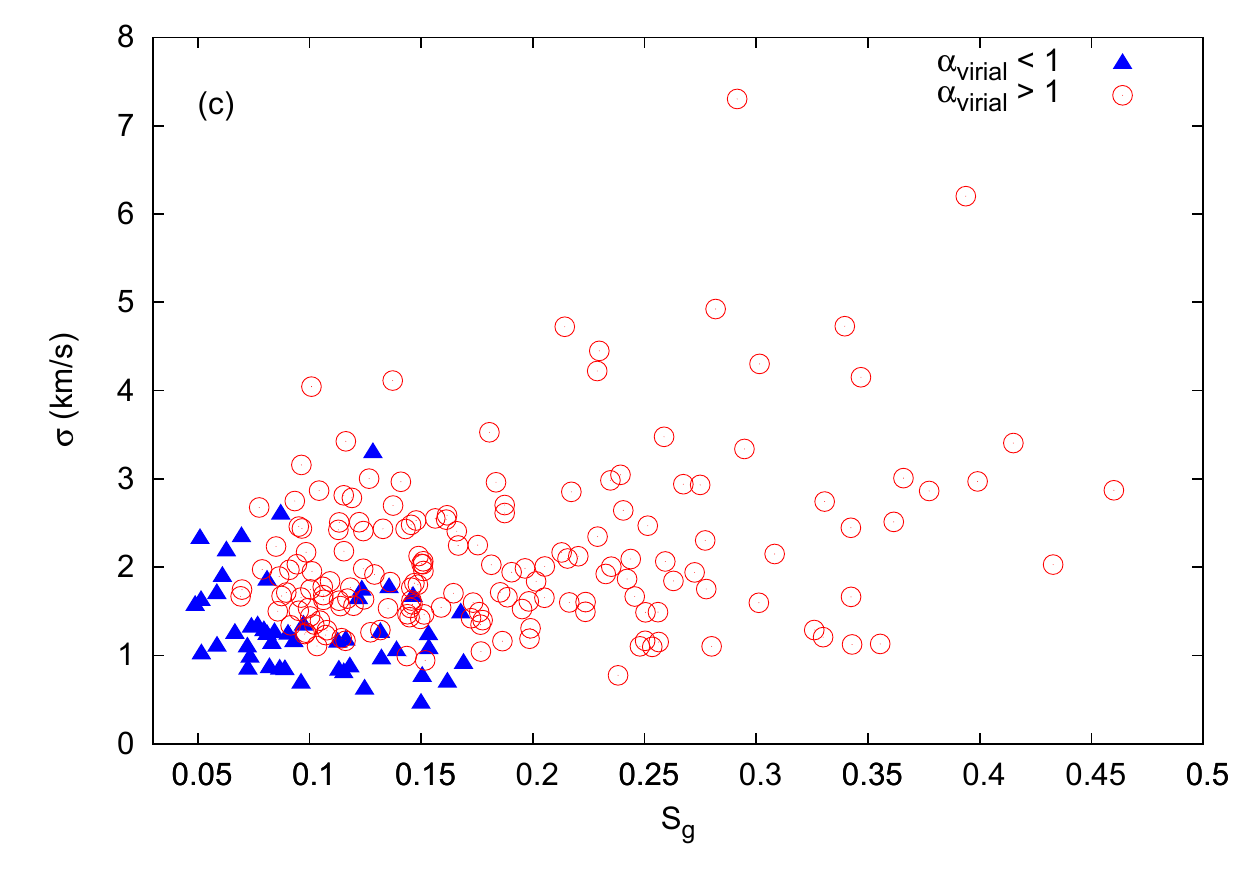}} 
\subfloat{\includegraphics[width=90mm,height=67mm]{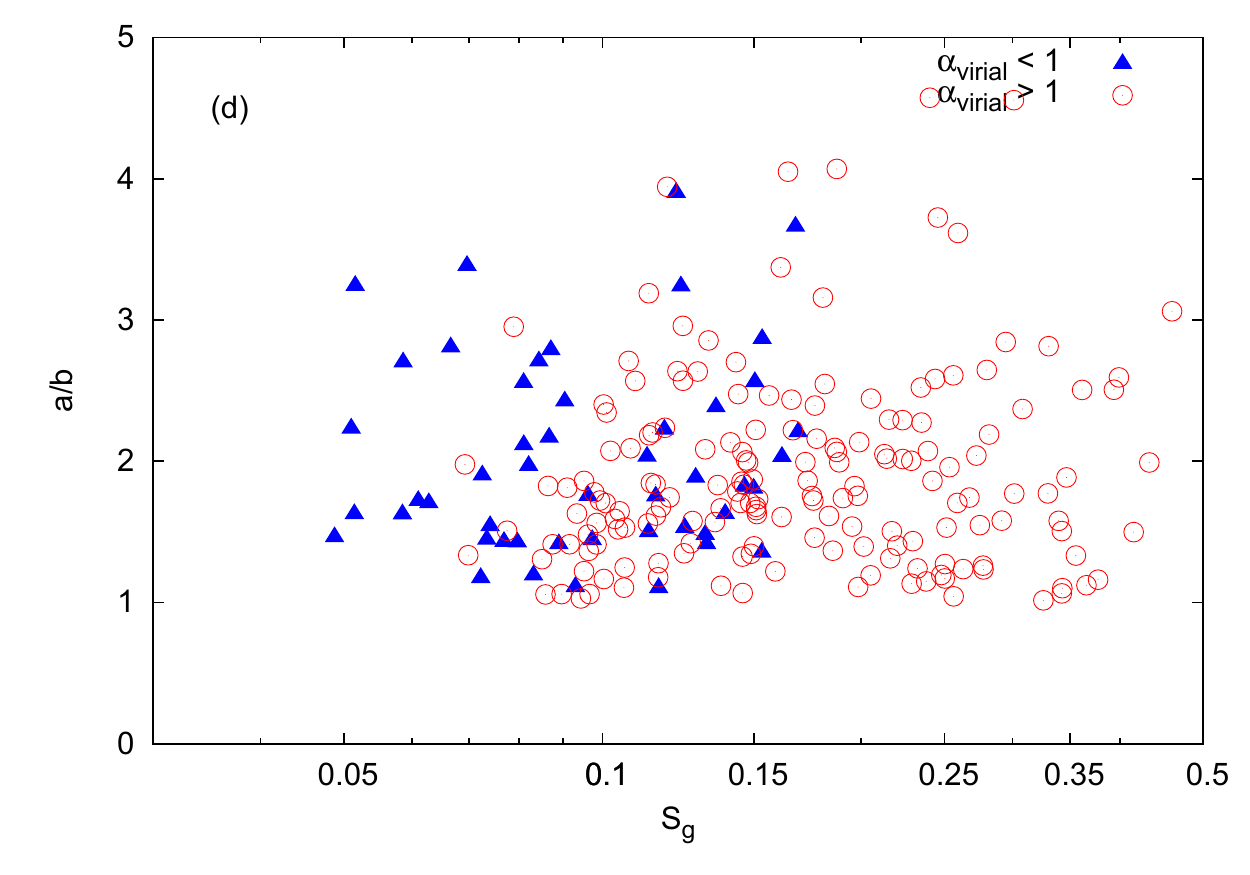}}                       
\caption{Relationship between the shear parameter $S_{g}$ and (a) the cloud mass $M_{\rm CO}$ in solar masses, (b) the cloud radius $r$ in pc, (c) the velocity dispersion $\sigma$ in km/s, and (d) the major-to-minor axes ratio, a/b. The blue triangles are the gravitationally bound clouds ($\alpha_{vir}<1$) and the red dots are the unbound clouds ($\alpha_{vir}>1$).}   
\label{fig:4}
\end{center}
\end{figure*}
We next compare the importance of shear for each cloud with its intrinsic properties such as mass, radius, velocity dispersion and major-to-minor axes ratio in Figure \ref{fig:4}. No strong correlations between GMC properties and shear has been found. Figure~\ref{fig:4}b shows a weak trend for the largest clouds to have higher $S_{g}$, suggesting that large clouds are more influenced by shear, although their $S_{g}$ values are all a factor of $>2$ less than unity and hence well below the threshold for disruption by shear (see Figure~\ref{fig:2}). However, Figure~\ref{fig:4}a shows no correlation between mass $M_{\rm CO}$ and $S_{g}$ values. Figure~\ref{fig:4}c suggests that the shear parameter, $S_{g}$, increases with increasing values of the velocity dispersion $\sigma$ but, as noted previously, this trend may reflect the covariance of $S_{g}$ with $\sigma$. \\

Dobbs $\&$ Pringle (2013) ran numerical simulations to test cloud evolution on large galactic scales. From their study of one particular cloud, they found that shear plays a key role in the dispersal of the cloud through its morphology, finding that filamentary clouds are more easily torn apart than round clouds, and that the filamentary morphology of GMCs is partially the result of shearing forces. The galaxy simulated by~Dobbs $\&$ Pringle (2013) is a massive spiral disk galaxy with a 2-armed spiral perturbation. Its stellar potential is deeper and the shear strength is correspondingly higher than in the LMC, which is a low-mass, irregular system. Figure 4d shows that there is no correlation between the major-to-minor axis ratio of the LMC GMCs and the shear parameter, and the elongated clouds exhibit a broad range of shear parameter values. One possibility is that local dynamical events play a more important role in shaping GMC morphology in the LMC than in disk galaxies, where shear is more dominant.\\

\begin{figure*}
\begin{center}
\includegraphics[width=100mm,height=75mm]{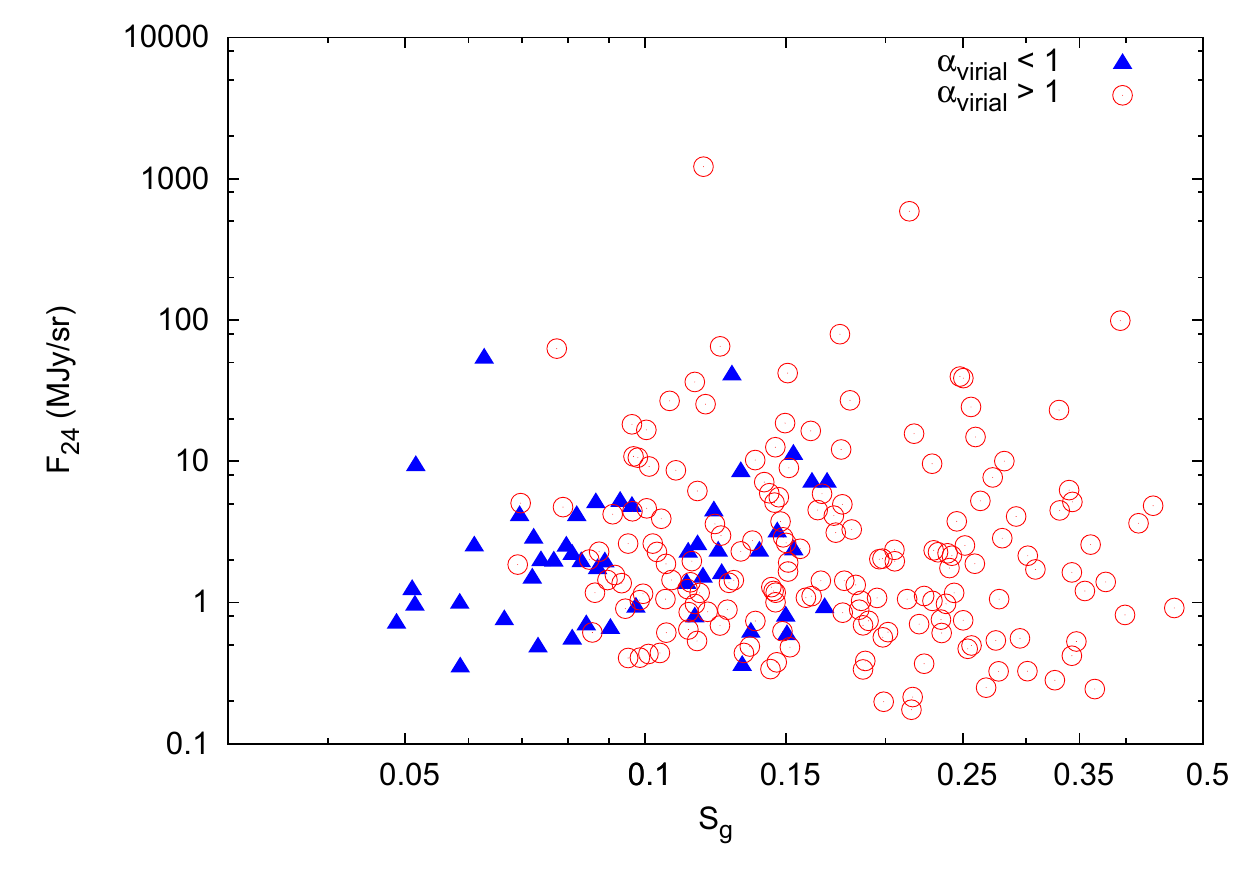}
\caption{Relationship between the shear parameter, $S_{g}$, for our selected sample of 260 resolved GMCs and the average 24~$\mu$m flux density, $F_{24}$, used as a star formation tracer. The blue triangles are the gravitational bound clouds ($\alpha_{vir}<1$) and the red dots are the unbound clouds ($\alpha_{vir}>1$).}
\label{fig:5}
\end{center}
\end{figure*}
Figure \ref{fig:5} shows the 24~$\mu$m flux density versus the shear parameter for each cloud. As previously discussed, the 24~$\mu$m flux can be used to infer star formation activity. Shear instability resulting in cloud disruption is expected to be an obstacle for star formation. We find no correlation between the shear parameter of the clouds and their star formation activity, which is not surprising given the low absolute values of $S_{g}$ for LMC GMCs and suggests that local and/or internal physical processes such as magnetic fields, turbulence and stellar feedback are more important for regulating the onset and progress of SF in LMC molecular clouds. 

%%%%%%%%%%%%%%%%%
\subsection{The tidal acceleration}
\begin{figure*}
  \centering
  \includegraphics[width=100mm,height=75mm]{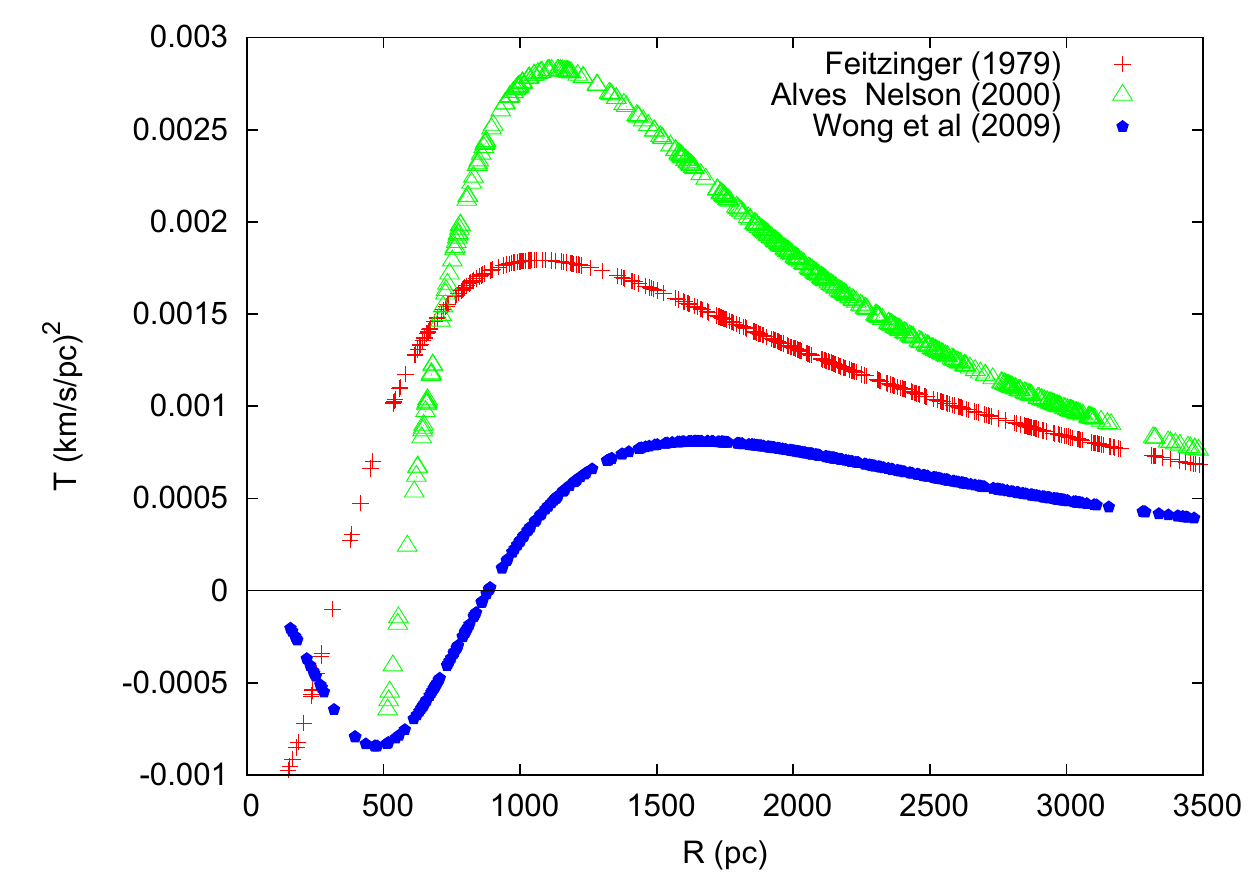}
  \caption{The tidal acceleration $T$ versus galactocentric radius $R$, where the $T$ values have been derived from the three different rotation curves of Feitzinger (1979), Alves $\&$ Nelson (2000) and Wong et al. (2009).}
  \label{fig:6}
\end{figure*}

%The range of tidal acceleration T values :
We plot the tidal acceleration $T$ as a function of the galactocentric radius $R$ for each rotation curve discussed in Section 3.3 in the Figure \ref{fig:6}. If we first consider the region of interest between $1000-3500$~pc, the tidal acceleration is in the range $500-2000$~km s$^{-2} \rm kpc^{-2}$, which are similar to the values derived by Blitz $\&$ Glassgold (1982) for the LMC. Logically, one expects that the tidal acceleration will decrease as we move away from the centre of the galaxy.
However we can see from Figure \ref{fig:6} that the tidal acceleration can take negative values in the inner 300~pc for the rotation curve from~ Feitzinger (1979), 500~pc for the rotation curve from~ Alves $\&$ Nelson (2000) and 1000~pc for~Wong et al. (2009). Those surprising values can be attributed to the fact that the galactic bulge remains poorly constrained, and indeed there are few data points in the inner part of the LMC (Figure \ref{fig:1}) to accurately constrain the rotation curve. Moreover the exact location of the kinematic galactic centre remains unknown. Beyond the inner few 100~pc the $T$ curves are more typical of the expected shape, so we need to be cautious of the results for the GMCs close to the galactic centre. As a precaution, while using the rotation curve from~Feitzinger (1979), we discard all the clouds located interior to a galactocentric radius of 300~pc, reducing our sample from 260 to 224 resolved clouds.\\

%%%%%%%%%%%%%
\subsection{Evaluation of the minimal mass from the Roche criteria}
With the $T$ values and the geometric and physical parameters of the GMCs, such as velocity dispersion $\sigma$, radius $r$ and CO mass $M_{\rm CO}$ from the catalog of ~Wong et al. (2011), we can now evaluate the Roche criteria using equation (\ref{eq:10}) for the 224 resolved GMCs exterior to 300~pc.\\

Once the Roche criteria is evaluated for each GMC, two pieces of information can be extracted. Firstly the sign of the Roche criteria tells us whether the GMC is tidally stable or not: if the expression corresponding to equation (\ref{eq:10}) is negative, then the total energy of the GMC is not sufficient to make the GMC reach the inner Lagrange point and the GMC is tidally stable. Secondly, by setting equation (\ref{eq:10}) to zero, the minimal mass, $M_{\rm min}$, of a tidally stable GMC can be calculated.\\

%Distribution of the minimal mass over the sample:
\begin{figure*}
\begin{center}
\includegraphics[width=100mm,height=75mm]{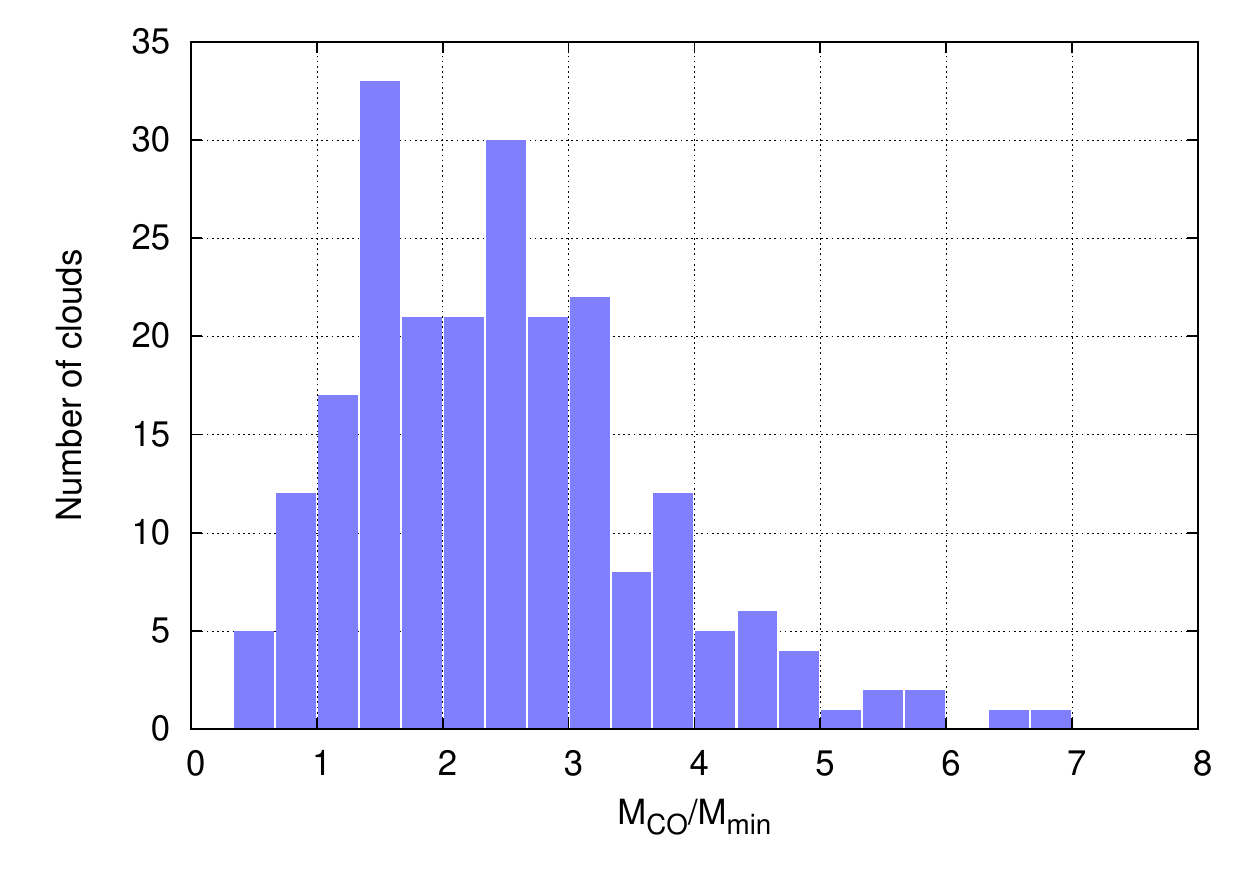}
\caption{Distribution of the ratio between observed cloud mass and minimal mass, $M_{\rm CO}/M_{\rm min}$, required for tidal stability for our selected sample of 224 resolved GMCs exterior to 300~pc.}
\label{fig:7}
\end{center}
\end{figure*}
Figure \ref{fig:7} shows the distribution of the ratio of observed cloud mass to the minimal mass,  $M_{\rm CO}/M_{\rm min}$, for our sample. The distribution ranges widely between 0.5 and 7, peaking around 2. However, very few of the GMCs appear to be tidally unstable, i.e. with $M_{\rm CO}$/$M_{min}<1$, and $80\%$ of our GMCs have $M_{\rm CO}$/$M_{min}$ between $1.5-4$. We thus conclude that the GMCs in the LMC are globally stable or at the edge of instability.\\

%Distribution of mass ratio accross the galaxy
\begin{figure*}
\begin{center}
\includegraphics[width=100mm,height=75mm]{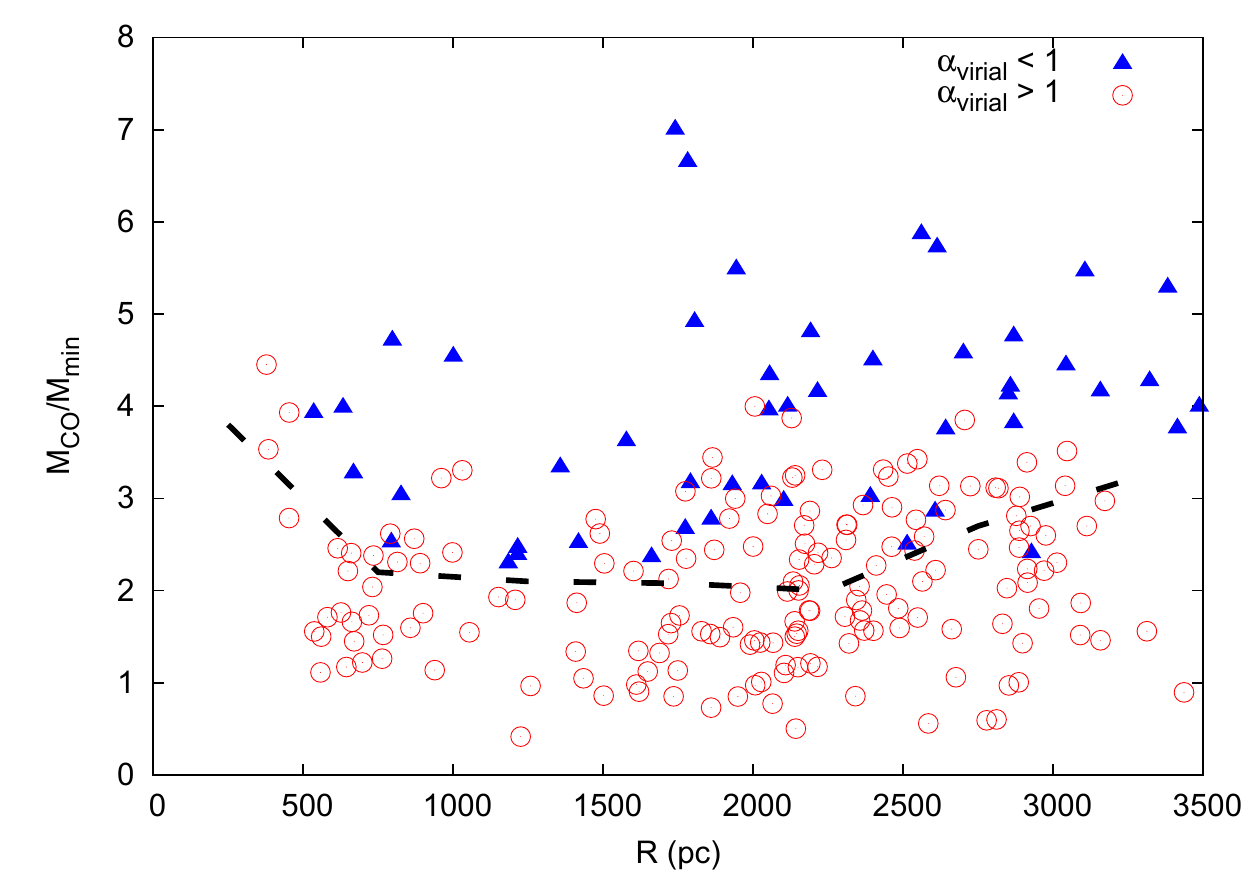}
\caption{Variation of the ratio $M_{\rm CO}$/$M_{\rm min}$ as a function of the galactocentric radius $R$ for our selected sample of 224 resolved GMCs exterior to 300~pc. The blue triangles are the gravitational bound clouds ($\alpha_{vir}<1$) and the red dots are the unbound clouds ($\alpha_{vir}>1$). The dotted line shows the median value of $M_{\rm CO}$/$M_{\rm min}$ for the entire sample of clouds in radial bins of 0.5 kpc.}
\label{fig:8}
\end{center}
\end{figure*}
Stark $\&$ Blitz (1978) suggested that because the tidal acceleration $T$ is supposed to be stronger in the inner region of the galaxy, GMCs closer to the galactic centre have to be denser in order to resist the stronger tidal stress. While the tidal acceleration $T$ takes negative values interior 300~pc (see Figure \ref{fig:6}), we can investigate a link between the position of the GMCs in the LMC and their $M_{\rm CO}$/$M_{\rm min}$ ratio beyond 300~pc. Figure \ref{fig:8} presents the variation of the mass ratio as a function of galactocentric radius. There is no specific location  with higher or lower tidal stability levels, and indeed the mass ratio ranges widely across all galactocentric radii. However we can remark that the sub-sample of unbound clouds ($\alpha_{vir}>1$) tend to have lower $M_{\rm CO}$/$M_{\rm min}$ ($<4$) than the gravitationally bound clouds ($\alpha_{vir}<1$), suggesting that velocity dispersion may be one of the most significant factors affecting the stability. One may wonder if the $M_{\rm min}$ for tidal stability has any variation with galactocentric radius.  If it does, then the lack of dependence of $M_{\rm CO}$/$M_{\rm min}$ on radius would be quite interesting, and might support the idea that cloud mass is limited by tides. However we also find no dependence of $M_{\rm min}$ on galactic radius. In the 30 Doradus region, clouds exhibit a mass ratio $M_{\rm CO}/M_{\rm min}$ ranging from $1.2-3.75$ with a mean value of $2.17$, which is consistent with the mean value over the entire sample $M_{\rm CO}/M_{\rm min}~\sim~2$. Thus both stability parameters ($M_{\rm CO}/M_{\rm min}$ and $S_{g}$) for clouds around 30 Doradus  do not show any significant difference compared to the values obtained for the complete sample of clouds.\\

%None correlation plots between Minimal mass and SF, Peak CO brightness, shapes etc...:
\begin{figure*}
\begin{center}
\subfloat{\includegraphics[width=90mm,height=67mm]{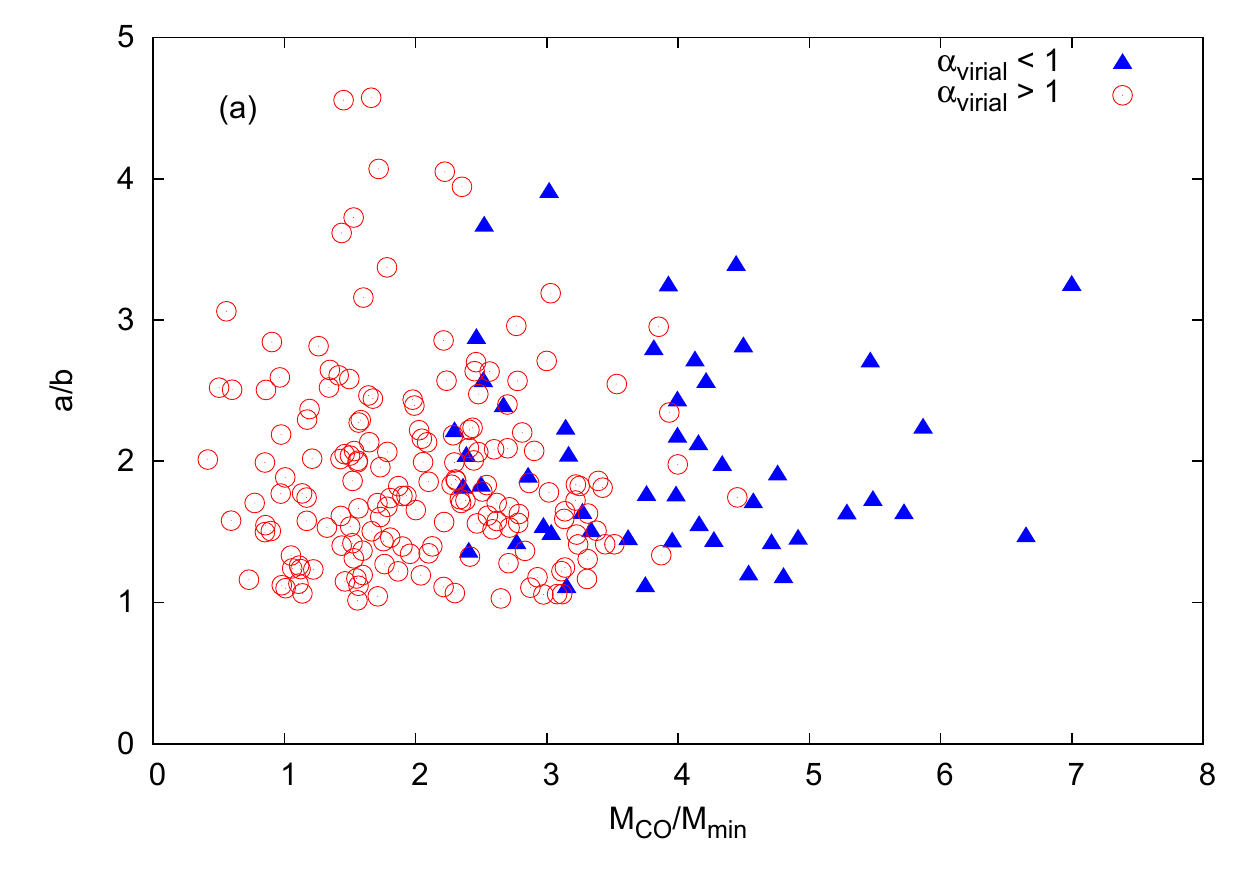}}                
\subfloat{\includegraphics[width=90mm,height=67mm]{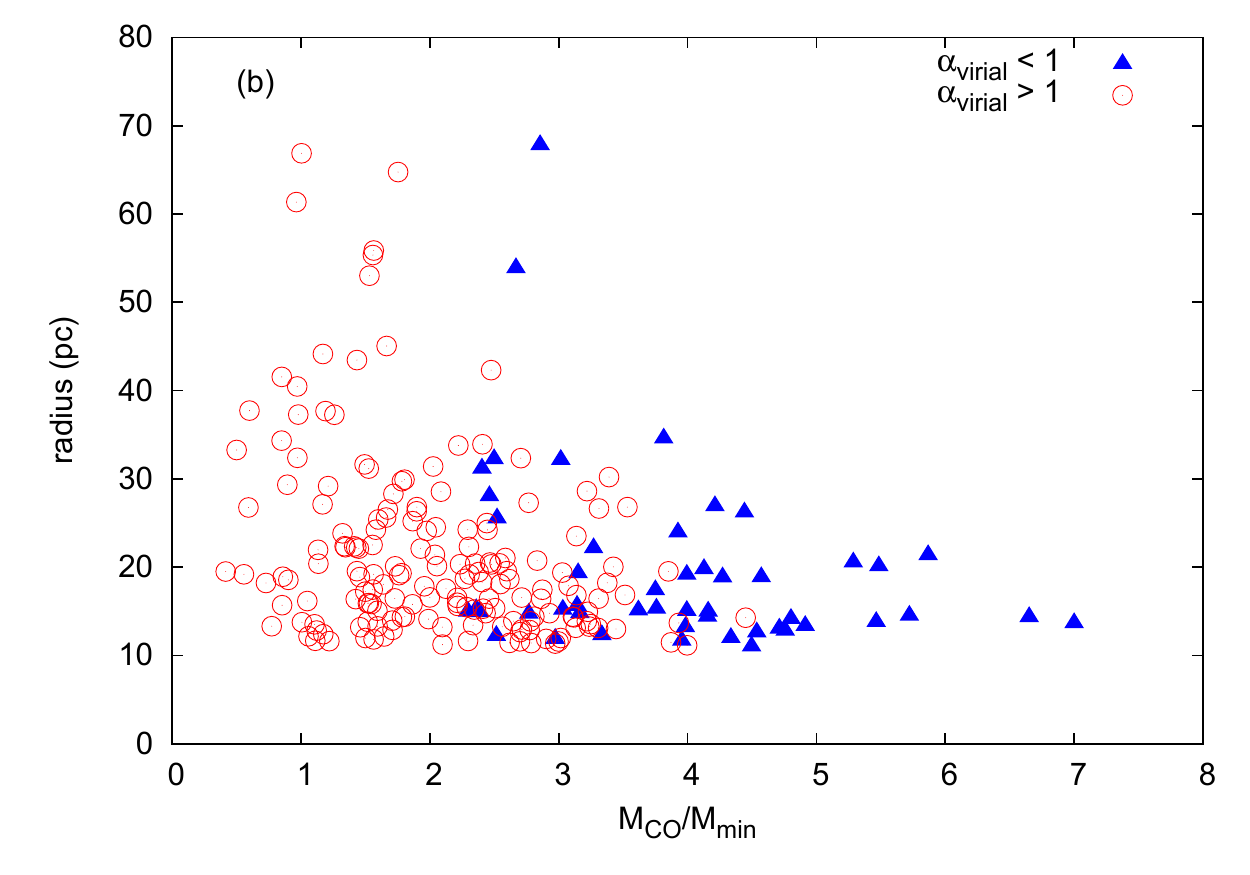}}\\     
\subfloat{\includegraphics[width=90mm,height=67mm]{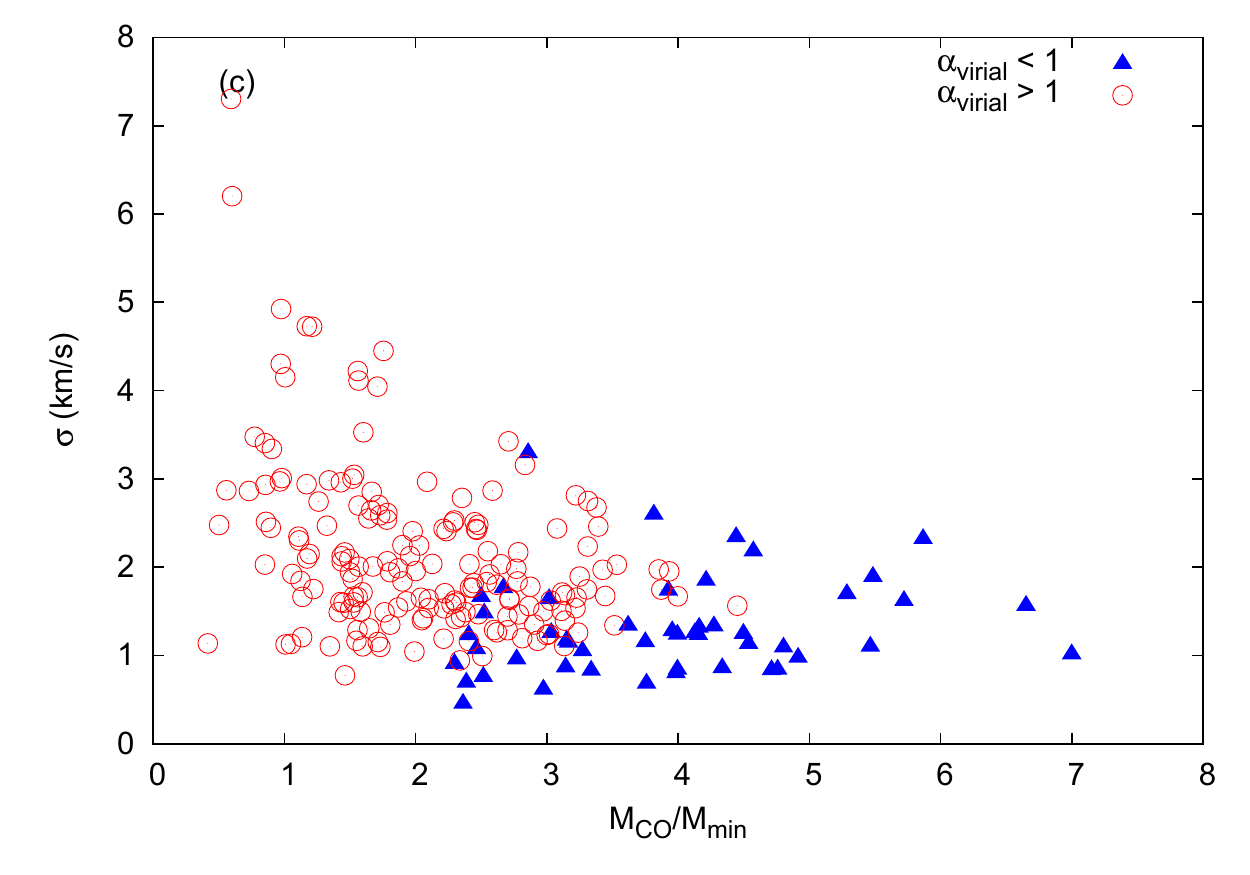}}                      
\caption{Relationship between the ratio $M_{\rm CO}$/$M_{min}$ and (a) the major-to-minor cloud axes ratio a/b, (b) the cloud radius $r$, and (c) the velocity dispersion $\sigma$. The blue triangles are the gravitationally bound clouds ($\alpha_{vir}< 1$) and the red dots are the unbound clouds ($\alpha_{vir} >1$).}   
\label{fig:9}
\end{center}
\end{figure*}
We next compare the values of $M_{\rm CO}$/$M_{min}$ of each cloud with its intrinsic properties such as major-to-minor axes ratio, radius, and value of the peak CO brightness temperature in Figure~\ref{fig:9}. Regarding the link between morphology and level of tidal stability, it may be worth noting in Figure \ref{fig:9}a that if a/b is $>$ 3.5, corresponding to an elongated cloud, then $M_{\rm CO}$/$M_{\rm min} <3$. Even if this trend between cloud morphology and mass ratio is not strong, it does suggest that filamentary clouds may be easier to tidally disrupt. Similarly, Figure \ref{fig:9}b also suggests that the most tidally stable clouds with $M_{\rm CO}$/$M_{\rm min} >  3$ tend to have smaller radii ($<$ 30 pc), whereas larger radii clouds ($>$ 40 pc) tend to have lower $M_{\rm CO}$/$M_{\rm min} < 3$. Even if the trend between small cloud radii and stability was noted in the shear parameter study, we stress the fact that, regarding the tidal stability, this trend involve a relatively small number of clouds ($<$ 27$\%$) in our sample. On the other hand, Figure \ref{fig:9}c shows that the $M_{\rm CO}$/$M_{\rm min}$ ratio decreases with increasing velocity dispersion, $\sigma$. This suggests that, as the velocity dispersion increases, the GMCs become more susceptible to disruption by tides. However, if GMCs follow a size-linewidth relation, then similar trends of $r$ and $\sigma$ with $M_{\rm CO}$/$M_{\rm min}$ should be expected. It is not clear from the correlation plots alone as to which is fundamental (i.e. from these plots, we can not distinguish whether large clouds tend to be closer to the tidal stability limit due to their size or their internal motions.)\\
\begin{figure*}
\begin{center}
\includegraphics[width=100mm,height=75mm]{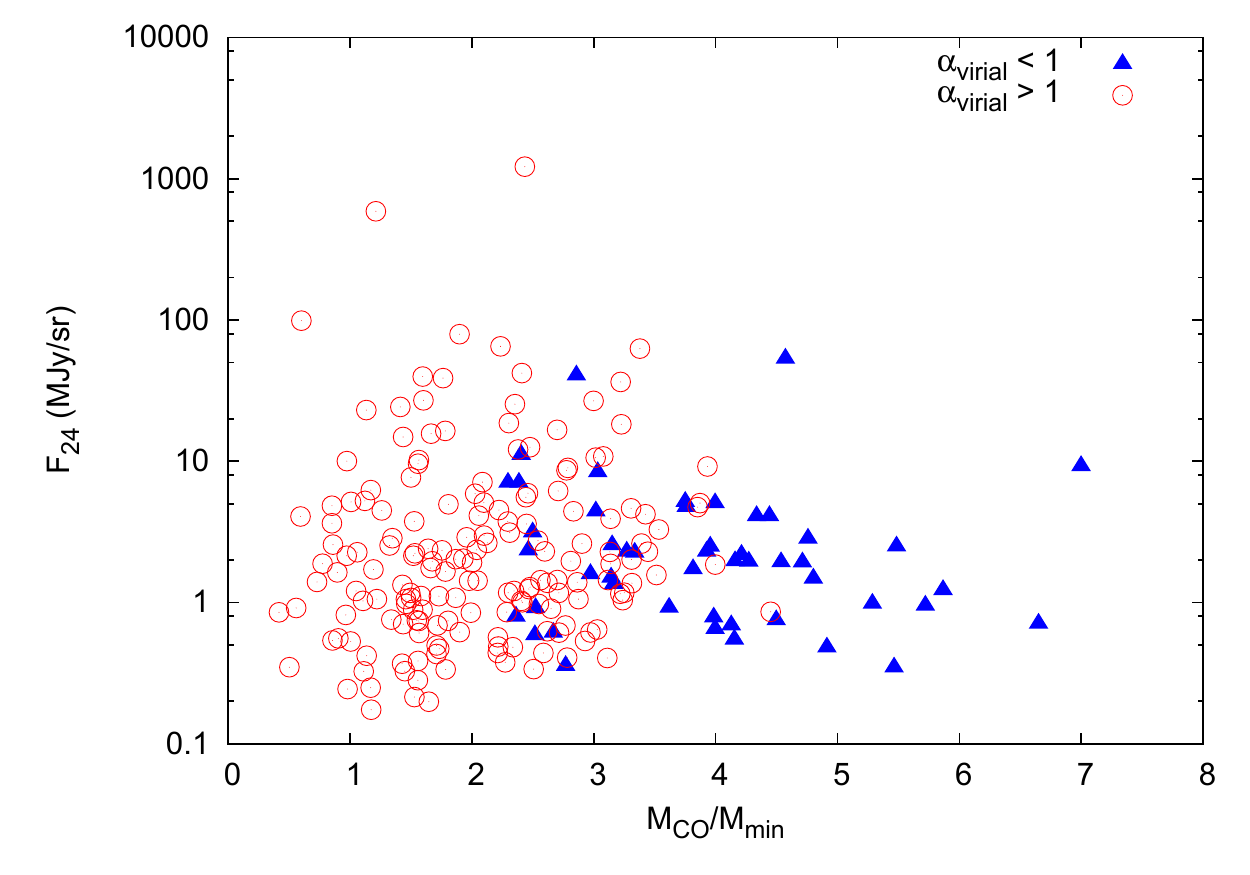}
\caption{Relationship between the ratio of the observed cloud mass to minimal mass, $M_{\rm CO}$/$M_{\rm min}$, and the 24~$\mu$m flux density used as a star formation tracer for our selected sample of 224 GMCs. The blue triangles are the gravitational bound clouds ($\alpha_{vir} < 1$) and the red dots are the unbound clouds ($\alpha_{vir} > 1$).}
\label{fig:10}
\end{center}
\end{figure*}

Finally, we check if the tidal interaction between the LMC and the GMCs affects the star formation activity. To test this hypothesis, we investigate whether clouds with lower $M_{\rm CO}$/$M_{\rm min}$ ratio (and hence more sensitive to tidal disruption) are the least active in terms of star formation. Figure \ref{fig:10} shows the 24~$\mu$m flux density versus the mass ratio for each cloud, and shows no correlation. Hence, there appears to be no relationship between a cloud's tidal stability and its star formation activity as determined by the 24~$\mu$m flux density.

%%%%%%%%%%%%%%%%%%%%%%%%%%%%%%%%%%%%%%%%%%%%%%%%%%%%%%%%%%%%%%%%%
\section{Discussion}
Galactic disk simulations run by~Dobbs $\&$ Pringle (2013) demonstrate that shear is a dominant mechanism for cloud disruption (stellar feedback and the general unbound state of GMCs also contribute), and that the influence of shear on GMCs is evident in their morphology, which become more elongated as the cloud is sheared and stretched. We examined whether there is evidence for this effect in the LMC by plotting the shear parameter versus the major-to-minor axis ratio for the clouds (Figure \ref{fig:4}d), but found no strong trends. We did, however, find that GMCs with larger radii have higher shear parameters. Indeed, larger GMCs have larger line widths (for which the velocity dispersion $\sigma$ is a proxy), which naturally causes the shear parameter to increase (since the velocity dispersion appears in the numerator of the shear parameter, see equation \ref{eq:6}). 
Overall, the discrepancy between our results and the predictions of ~Dobbs $\&$ Pringle (2013) tends to support the idea that different physical processes are responsible for cloud disruption in different galactic environments and, as a corollary, that GMCs may have characteristic lifetimes that also vary with environments. We note, however, that ~Dib et al. (2012) also found that shear plays only a minor role in the evolution and star formation activity of molecular clouds in the Milky Way (which should be more akin to the system simulated by ~Dobbs $\&$ Pringle (2013) than the LMC.)\\

Yang et al. (2007) derived a multicomponent Toomre criterion for the LMC, finding that $85\%$ of their massive YSO candidates lie in gravitationally unstable regions, implying that star formation occurs predominantly in these regions. They conclude that large-scale gravitational instability is the ultimate driver of star formation. Our finding that shear and galactic tides do not exert a strong influence on the star-forming activity within GMCs is not necessarily in contradiction to this, since large-scale gravitational instability may regulate the formation of GMCs, even if star formation on cloud-scales is independent of large-scale dynamical effects once a GMC has been assembled. We further note that Hunter et al. (1998) used the Toomre criterion and the shear parameter to study the gravitational state of GMCs in 15 irregular galaxies and suggested that the Toomre criterion was overestimating the instability. Using $A$ to quantify the gas kinematics instead of the traditional epicyclic frequency $\kappa$ for the Toomre criterion may produce a difference in the instability threshold, especially for rising rotation curves and low shear environments like the LMC. As a result, the estimated shear parameter $S_{g}$ can be significantly lower than the Toomre parameter $Q_{g}$. Hunter et al. (1998) concluded that even if the shear model was better able to reproduce the observations than the Toomre model, both models fail to probe star formation activity, suggesting that other processes are important in irregular galaxies.\\ 

Consider now the stability of GMCs against galactic tidal forces, an idea first introduced by Stark $\&$ Blitz (1978), which assesses the balance between cloud's self-gravity, internal (thermal + turbulent) pressure and the large-scale galactic tide. They found that the Milky Way's GMCs are only 3 times more massive than the tidal disruption limit, suggesting that tidal forces can have an influence on GMC morphology. In our study of the LMC, we indeed found that $\sim 70\%$ of our sample have $M_{\rm CO}/M_{min} <$ 3, and found a weak trend for highly elongated clouds to have a lower mass ratio. However, the weakness of this trend forces us to remain cautious with regards to the link between tidal effects and cloud's morphology. \\

Blitz (1985) studied a GMC sample in M31. Since tidal acceleration values and average masses of in that sample are comparable to those of~Stark $\&$ Blitz (1978), Blitz concluded that GMCs in M31 may also have their mass, size and morphology tidally limited. In our study of the LMC using the shear and tidal effects, we also noticed a tendency for clouds with larger radii ($> 35$ pc) to be more susceptible to tidal disruption. Indeed, larger GMCs are spatially more extended and thus are likely closer to filling their Roche lobes, which causes them to be more sensitive to tidal effects. Since $\sim 90\%$  of our clouds have a radius below that limit, one might suggest that there are not many clouds above that size because of stronger tidal effect.  \\

Rosolowsky $\&$ Blitz (2005) evaluated the tidal stability of GMCs in M64, and found significant shear in the inner part of the galaxy ($R<$ 400~pc), where clouds are marginally stable against tides. The shear parameter evaluation of the inner region of the dwarf galaxies NGC 2915 and NGC 1705 performed by ~Elson et al. (2012) shows a clear relation between instability and absence of star formation activity in that zone. A general result is that shear and tidal instability seem to play a more significant role in regulating GMC evolution and star formation in the inner regions of galactic disks.
Recently, Colombo et al (submitted) have shown that the GMC mass spectrum is strongly truncated in the nuclear bar region of M51 (at $R<1~$kpc), and propose that this truncation is evidence for shear limiting the growth of high-mass GMCs in this zone. In the LMC, by contrast, Wong et al. (2011) found no evidence for a truncation in the GMC mass spectrum -- though we note that the characteristic mass of GMCs in the LMC is already much lower than in M51 -- and our analysis in this paper indicates no systematic radial variation in the shear parameter or tidal stability of LMC clouds. In M33, the existence of a truncation in the GMC mass spectrum for $R<2.1~$kpc is still debated (cf Rosolowsky et al. (2007) and Gratier et al. (2012)). M33's rotation curve at these radii has a similar shape to that of the LMC, although the characteristic surface densities of gas and stars are somewhat greater. Taken together, these results again point to the importance of the galactic environment on GMC evolution. A more detailed study of the connection between the shape of the GMC  mass spectrum and the dynamical environments within galactic disks is clearly merited, and a topic that will benefit from ALMA's ability to survey the molecular gas in nearby galaxies across a range of Hubble types at cloud scale resolution.\\

While shear and tidal forces may be important for the formation of GMCs in the LMC, we conclude from our investigations that once a GMC is formed, both shear and tidal effects have a very little impact on GMC's stability. We find that both the shear parameter and the tidal acceleration maintain an almost constant mean value across the LMC. Moreover, GMCs potentially tidally unstable do not have lower star formation activity than the average, neither do GMCs subject to higher shear parameter value. Since we remark that clouds with large radii seem more affected by both tidal and shear effects, we suspect that either these effects may be noticeable only on larger scales than a typical resolved GMC, or tides may have limited the GMC sizes. Furthermore, tidally unstable clouds have lower virial parameter (with high velocity dispersions) than tidally stable clouds, suggesting that internal physical processes of the clouds may be key to their stability state.\\

Dib et al. (2012) evaluated the shear for a large sample of resolved GMCs in the Milky Way and found similar conclusions for almost all the molecular clouds: there is no evidence that shear is playing a significant role in instability or in star formation activity, and moreover the shear parameter of the clouds does not depend on their position in the Galaxy. These conclusions may be extended beyond the case of the LMC and the Milky Way in future studies.

%%%%%%%%%%%%%%
\subsection{The effect of $X_{\rm CO}$}
The value of the CO-to-H$_{2}$ conversion factor $X_{\rm CO}$ is supposed to be an inverse function of the metallicity (Narayanan et al. 2012), but remains poorly constrained. In this work we used $X_{\rm CO} = 3.0\times10^{20}$ cm$^{-2}$ (K km s$^{-1}$)$^{-1}$ derived by Leroy et al. (2008) for the LMC. Many studies (Rosolowsky $\&$ Leroy 2006; Wong et al. 2011) adopt the well established value for the Milky Way of $2.0\times10^{20}$ cm$^{-2}$ (K km s$^{-1}$)$^{-1}$, however since the metallicity of the LMC is slighly lower that the Milky Way's, the LMC's $X_{\rm CO}$ is likely higher by a factor of two, which corresponds to $X_{\rm CO} = 4.0\times10^{20}$ cm$^{-2}$ (K km s$^{-1}$)$^{-1}$ taken by~ Hughes
et al. (2010) when they derived their GMC catalogue. Thus the proposed value of Leroy et al. (2008) seems to be a good compromise for our study.\\

To assess the impact of the conversion factor on our results, we performed the tidal stability study using three different values of $X_{\rm CO}$. Adopting an $X_{\rm CO} = 2\times10^{20}$ cm$^{-2}$(K km s$^{-1}$)$^{-1}$ we find that $\sim20$\% of the GMCs are tidally unstable, whereas an $X_{\rm CO} = 4\times10^{20}$ cm$^{-2}$(K km s$^{-1}$)$^{-1}$ resulted in no unstable clouds. A summary is presented in the Table \ref{table:table2}. Nevertheless, regardless of the choice of $X_{\rm CO}$, tidal instability leading to disruption remains marginal through our GMCs sample. A better handle on the value $X_{\rm CO}$ in the LMC is important, but is beyond the scope of this work.\\

%%%%%%%%%%%%%
\subsection{Impact of rotation curve}
The results presented in Section 4 used the rotation curve of~Feitzinger (1979). In Appendix \ref{fig:Annexe1} and \ref{fig:Annexe2} we present results using different rotation curves. We plot the distribution of the shear parameter for the three rotation curves used in this study in Figure \ref{fig:Annexe1}. The rotation curve of ~Feitzinger (1979) results in shear parameter values from 0.03--0.48, with a peak at $S_{g} \sim$~0.12. Using the rotation curve from~ Wong et al. (2009), the distribution ranges between 0.02--0.35 and peaks at 0.08, while the rotation curve from~ Alves $\&$ Nelson (2000) results in $S_{g}$ values between 0.02--0.5 and peaks at 0.12. Globally the range and peak in the shear parameter for our cloud sample are similar regardless of which rotation curve is chosen, and our conclusions regarding cloud properties and stability remain unchanged.  \\

Regarding the tidal acceleration, Figure \ref{fig:Annexe2} presents the minimal mass $M_{\rm min}$ of the GMCs plotted against the cloud mass derived with $X_{\rm CO}$= $3.0\times10^{20}$ cm$^{-2}$ (K km s$^{-1}$)$^{-1}$  using the rotation curve of Feitzinger (1979), Wong et al. (2009) and Alves $\&$ Nelson (2000). These plots represent the tidal state of the GMCs: if they are below the 1-to-1 line, their mass is greater than the minimal mass and they are tidally stable. Figure \ref{fig:Annexe2} also shows that the GMCs in the LMC are globally tidally stable irrespective of the rotation curve used. Table \ref{table:table2} summarises the limited impact of both rotation curve and $X_{\rm CO}$ on our results using the tidal acceleration as instability factor.
\begin{table*}
\begin{center}
\caption{Percentage of tidally unstable clouds using various $X_{\rm CO}$ factors and rotation curves.} 
\begin{tabular}{l c c c}
\hline\hline
 Rotation curve & \multicolumn{3}{c}{$X_{\rm CO}$  (cm$^{-2}$ (K km s$^{-1}$)$^{-1}$)}  \\
 & $2.0\times10^{20} \, $  & $3.0\times10^{20} \, $ & $4.0\times10^{20} \, $ \\
 \hline
 Feitzinger (1979) & 25$\%$   &   5$\%$  &   0.0$\%$\\
 Wong et al. (2009) & 12$\%$   &   1.8 $\%$ &  0.0$\%$\\
 Alves $\&$ Nelson (2000) & 33$\%$ & 8$\%$  & 1$\%$\\
 \hline\hline
\label{table:table2}
\end{tabular}
\end{center}
\end{table*}

%%%%%%%%%%%%%%%%%%%%%%%%%%%%%%%%%%%%%%%%%%%%%%%%%%%%%%%%%%%%%%%%,
\section{Conclusions}
We have studied the stability of a sample of more than 200 resolved GMCs located in the Large Magellanic Cloud using the MAGMA $^{12}$CO~(J=1-0) survey (Wong et al. 2011). We examined the gravitational stability of the GMCs against galactic rotational shear and their tidal stability by evaluating the shear parameter following the method of Dib et al. (2012), and the tidal effect of the LMC using the method of Stark $\&$ Blitz (1978). We report the following results and conclusions: 
\begin{enumerate}
\item Galactic shear does not seem to be important regarding the stability of our GMC sample in the LMC. The distribution of the shear parameter $S_{g}$ peaks around 0.12, which is far from the domain where shear causes cloud disruption ($S_{g} > 1$). This result hold regardless of which rotation curve we use to calculate the Oort constant value $A$.
\item The GMCs in the LMC seem to be globally tidally stable. The distribution of the ratio between their actual mass and the minimum mass required for tidal stability, $M_{\rm CO}$/$M_{\rm min}$, peaks around 2. This result holds regardless of the $X_{\rm CO}$ value we use when calculating $M_{\rm CO}$ or which rotation curve we use for calculating the tidal acceleration $T$. This result is consistent with the absence of truncation at the upper end of the cloud mass distribution.
\item Both the shear parameter $S_{g}$ and ratio $M_{\rm CO}$/$M_{\rm min}$ show no systematic variation with distance from the center of the galaxy. 
\item No correlation was found between the 24~$\mu$m flux density (chosen as a star formation tracer) and stability against shear or tidal disruption. It appears that once the GMCs are formed, their star formation efficiency does not depend on the shear parameter or galactic tidal effects.
\item We find no obvious correlation between cloud properties such as galactic position, mass, axis ratio and peak CO brightness temperature and the GMC's stability against galactic scale dynamical effects. Nevertheless, a weak tendency for clouds with high velocity dispersion to be tidally unstable was found.  
\item A weak trend was found in which  GMCs with larger radii (r $>$ 40 pc) are less resistant against both shear and tidal instability. 
\end{enumerate}

The fact that similar conclusions were found using three different rotation curves covering a broad range of tidal acceleration and shear parameter values strengthens our conclusions. Our study, based on the work of Stark $\&$ Blitz (1978) and Dib et al. (2012) and applied to the GMC catalog of Wong et al. (2011), does seem to exclude galactic tides and shear as playing a key role in GMCs instability and star formation processes in the LMC. This result supports the conclusions of Dib et al. (2012). Thus, shear and tidal effects may be important on large scales in the more diffuse ISM (so therefore could be important for GMC formation), but that once the GMCs form, they achieve a dynamical configuration that is no longer susceptible to modifications by shear or tides. Analysing the GMCs at smaller scales is necessary to understand the evolution of the ISM and star formation processes inside GMCs.\\

Finally, we must note the limitations of our approach. The formula for the growth of a density perturbation against shear involves surface density $\Sigma$ and velocity dispersion $\sigma$, but it is not clear that we should be using the observed values for these quantities, since what we are observing may already be the result of gravitational amplification. In fact, equation (\ref{eq:4}) with three different $\Sigma$'s suggests that these analytic estimates are only a first step. Similarly, it is not obvious we should use observed values of velocity dispersion $\sigma$, mass $M$ and cloud radius $r$ in equation (\ref{eq:10}) to evaluate the Roche criterion, since those observed values may be obtained under galactic tidal influence. \\

Moreover it must be noted that we have not included the effects of magnetic fields in GMCs, which may help to temporally stabilise the GMC by confinement (Shu et al. 1987). We also did not consider the internal cloud motion. The GMCs from the catalog of Hughes et al. (2010) clearly have a velocity gradient, with the velocity dispersion varying by up to a factor of three from the cloud centre to the cloud edge. A more elaborate investigation of this phenomenon has been made by Ballesteros-Paredes et al. (1999) by studying the velocity and density structure inside molecular clouds. The LMC GMCs may in fact be very inhomogeneous: they are likely composed of many sub-structures which can be gravitationally unstable and enhance star formation, as suggested by Rodriıguez (2005).
\section*{Acknowledgements}
This work was partially supported by a Swinburne Faculty of ICT Research Grant Scheme. We thank the anonymous referee for helping improving the quality of the paper. AH thanks Sharon Meidt for useful discussions, and acknowledges funding from the Deutsche Forschungsgemeinschaft (DFG) via grant SCHI 536/5-1 and SCHI 536/7-1 as part of the priority program SPP 1573 'ISM-SPP: Physics of the Interstellar Medium'. 

\bibliographystyle{plain}

\begin{appendix}
\section{The effect of rotation curve}

\begin{figure*}
\begin{center}
  \subfloat{\includegraphics[width=60mm,height=45mm]{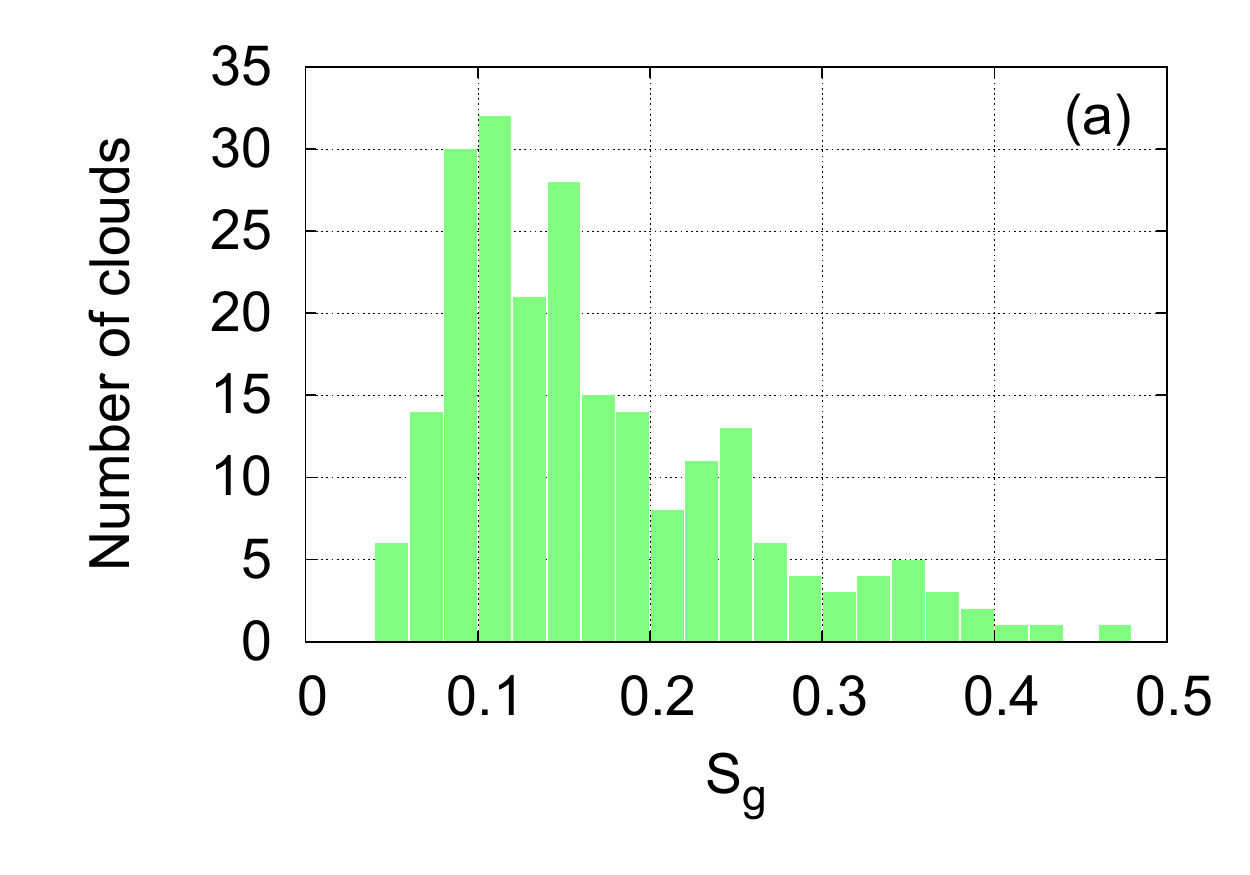}}                
  \subfloat{\includegraphics[width=60mm,height=45mm]{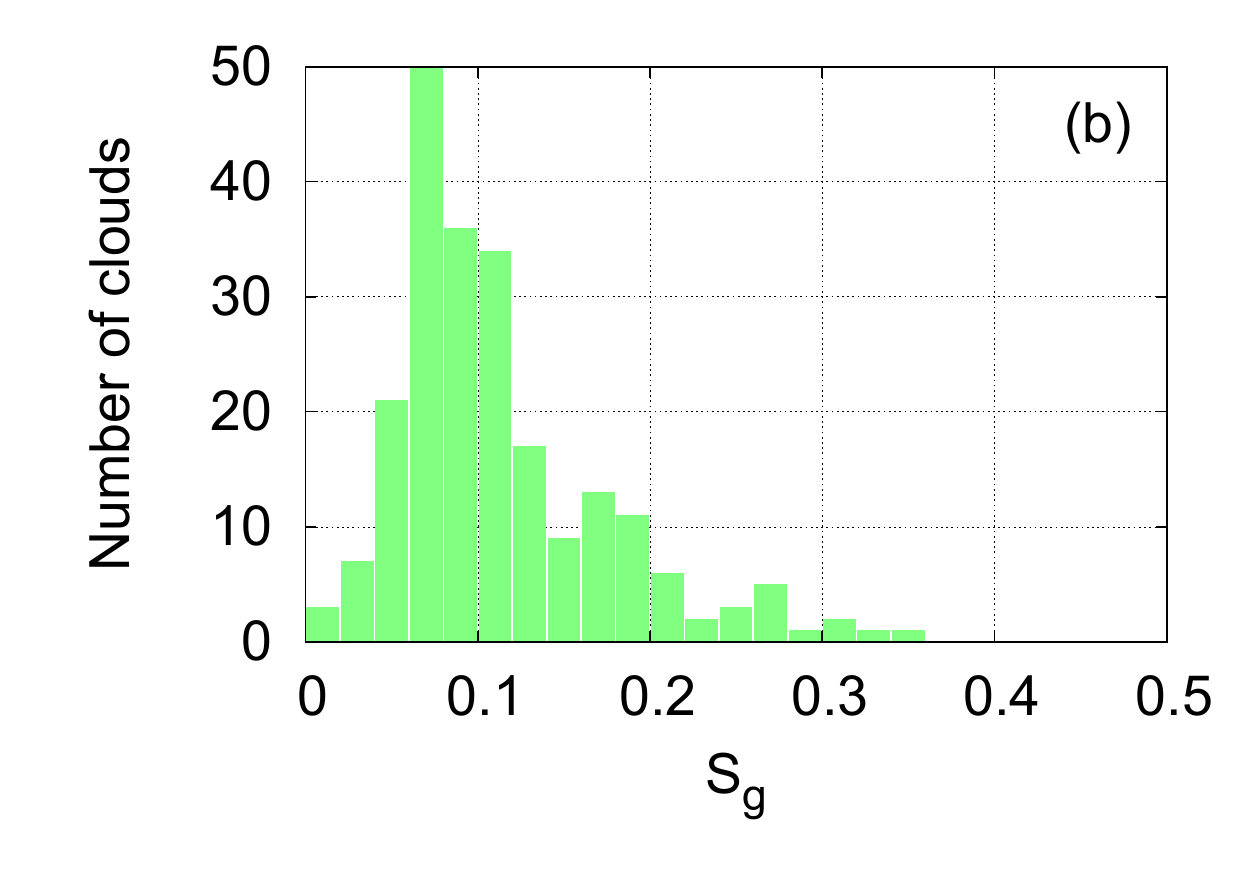}} 
  \subfloat{\includegraphics[width=60mm,height=45mm]{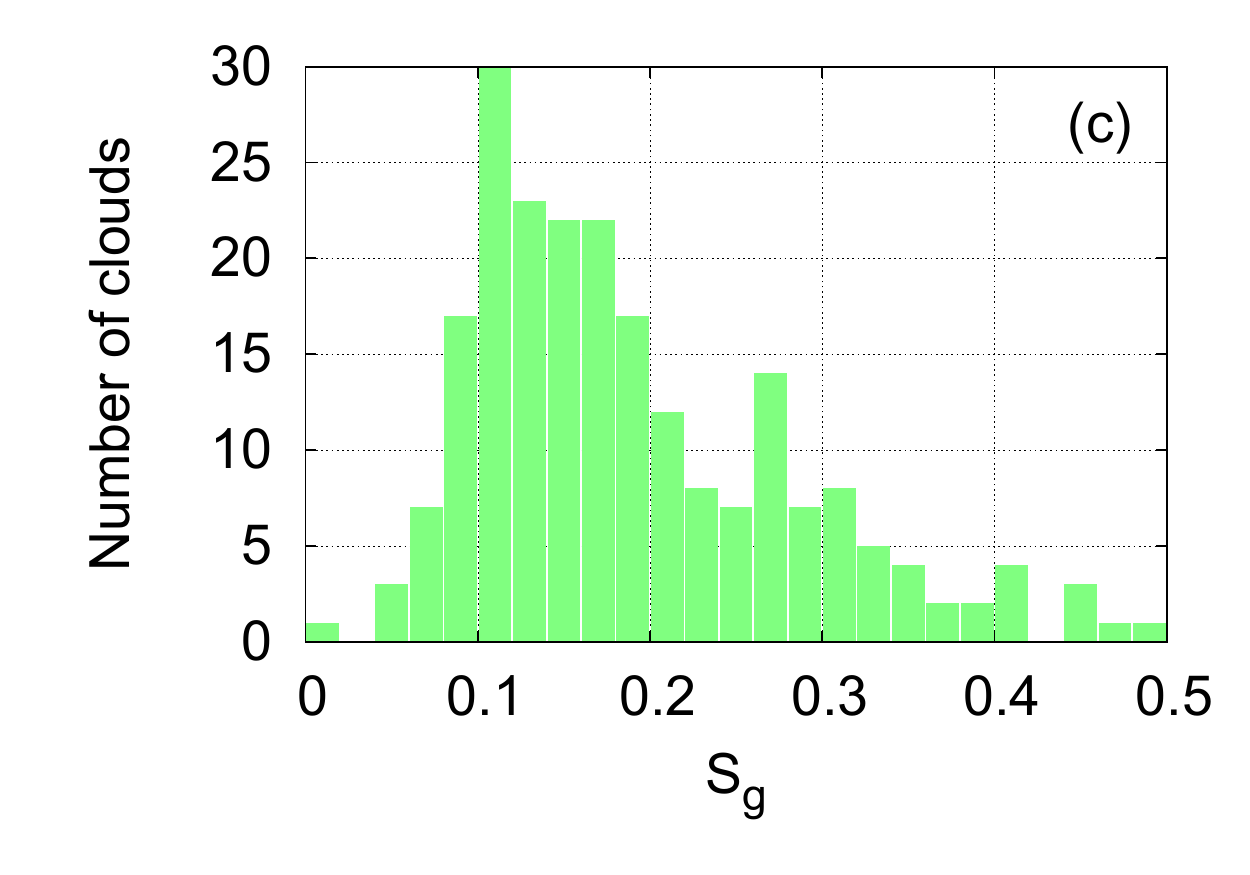}}     
  \caption{Distribution of the shear parameter $S_{g}$ for our selected sample of 260 resolved GMCs using the rotation curve of (a) Feitzinger (1979), (b) Wong et al. (2008), and (c) Alves $\&$ Nelson (2000). }   
  \label{fig:Annexe1}
\end{center}
\end{figure*}
 \begin{figure*}
\begin{center}
  \subfloat{\includegraphics[width=60mm,height=45mm]{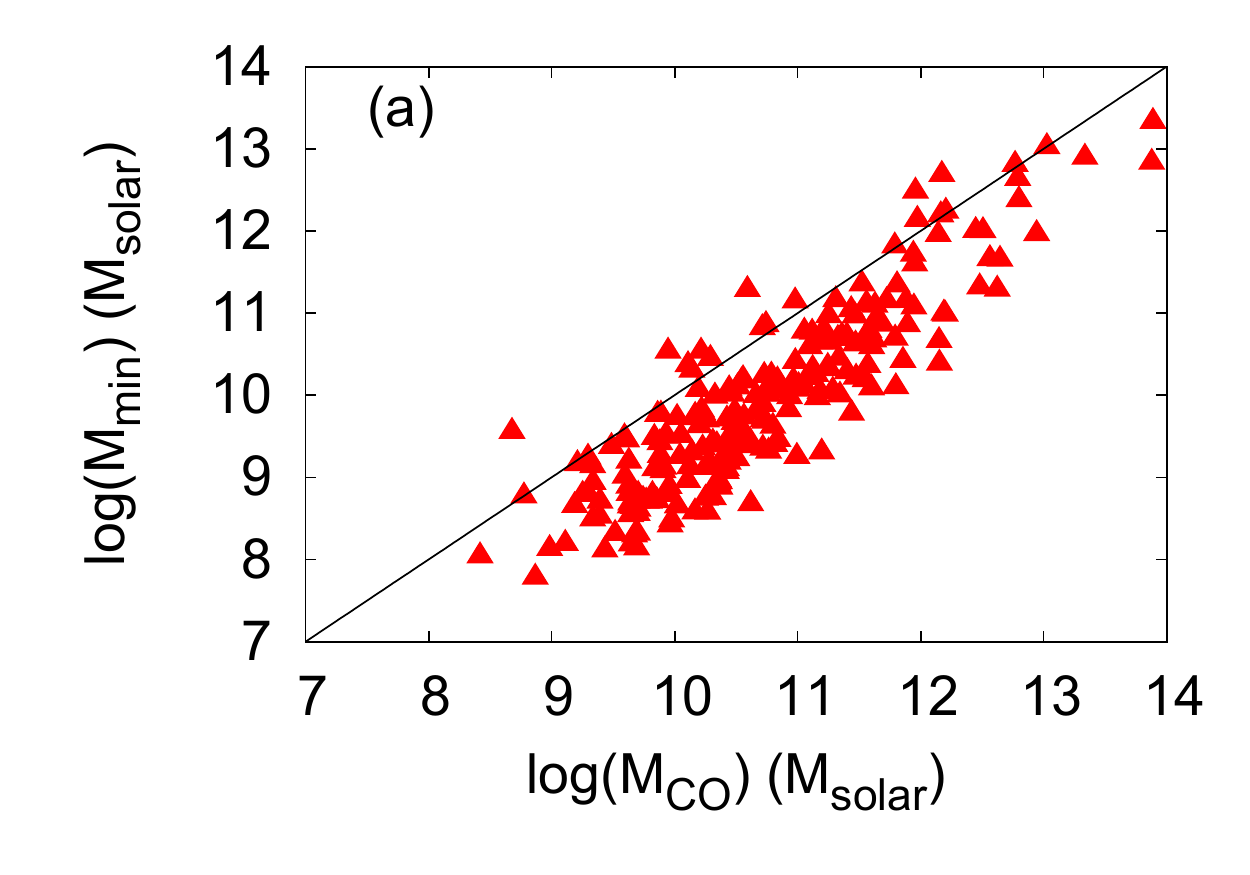}}                
  \subfloat{\includegraphics[width=60mm,height=45mm]{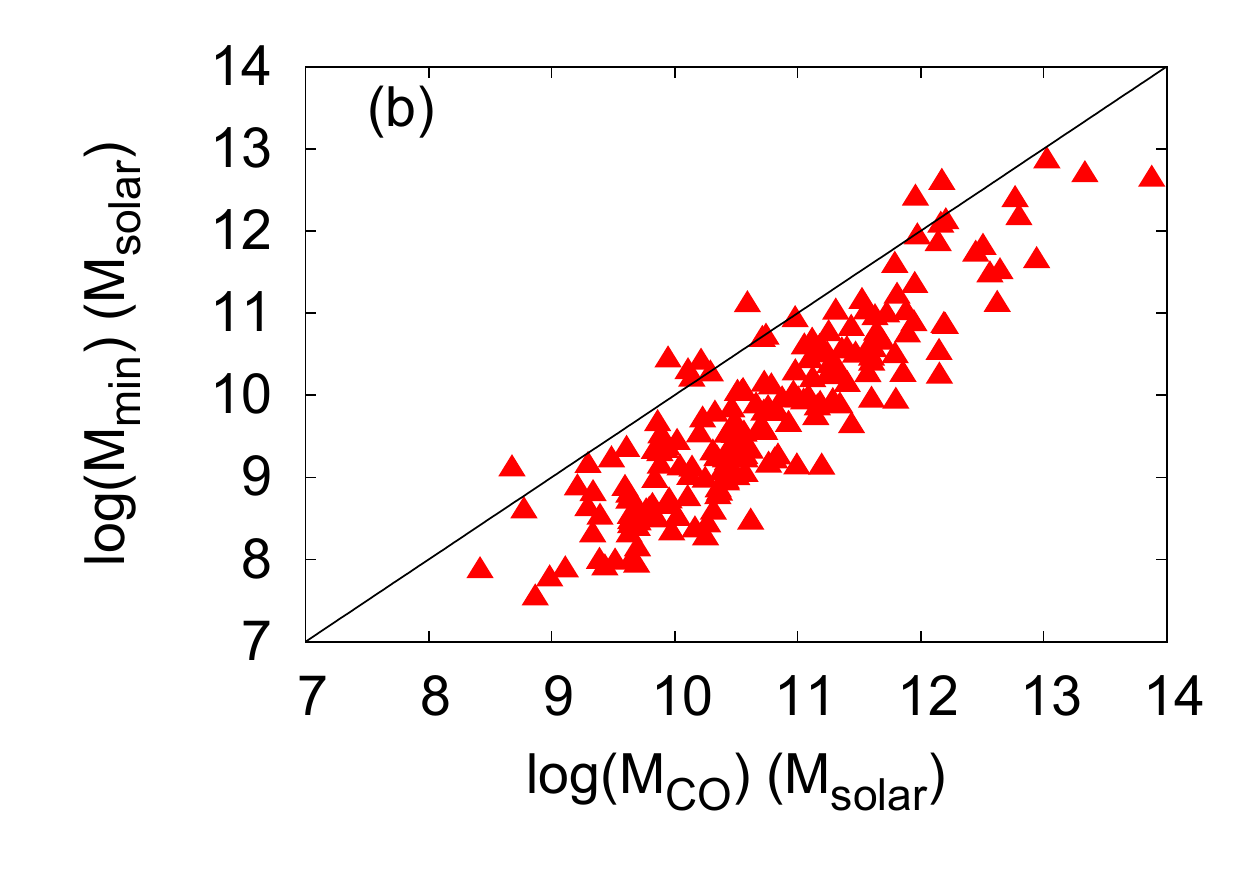}}  
  \subfloat{\includegraphics[width=60mm,height=45mm]{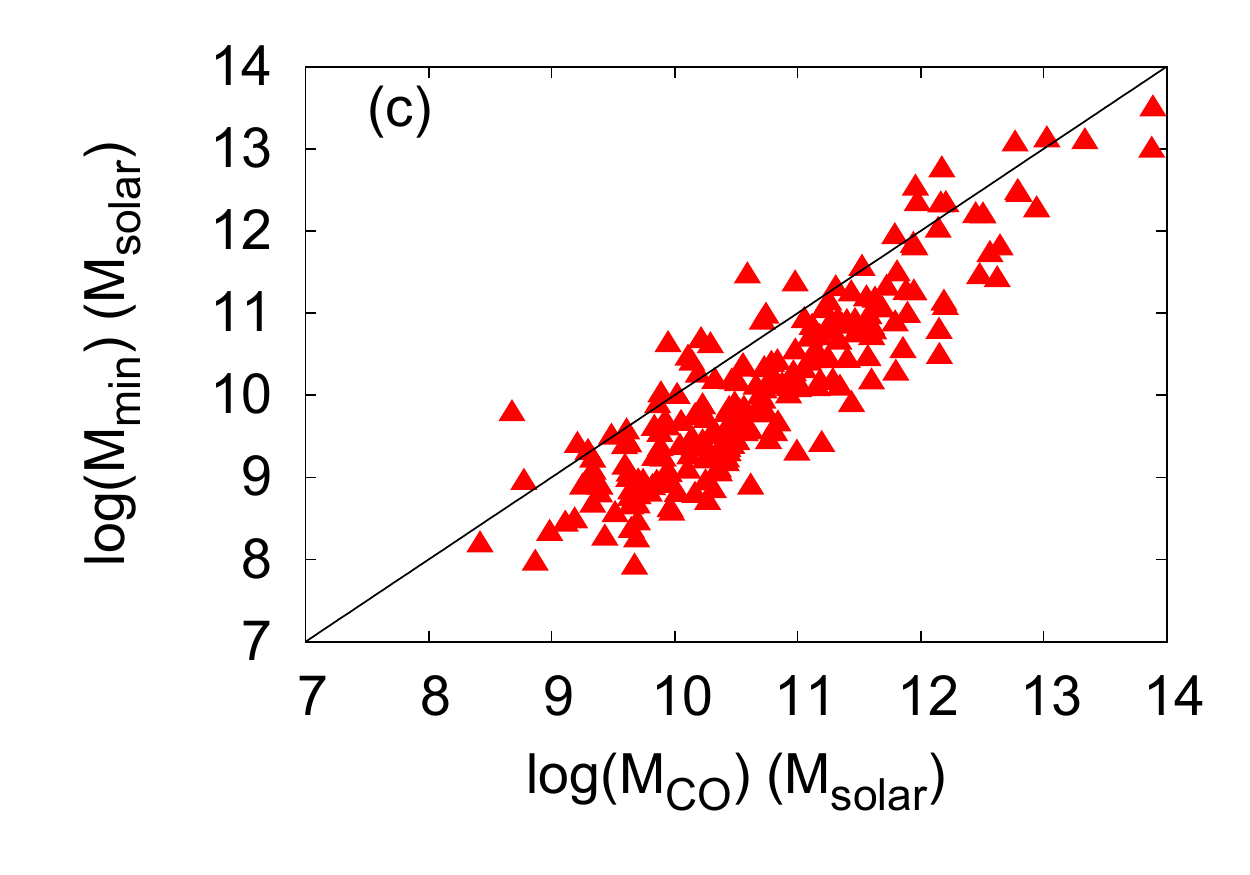}}            
  \caption{Observed cloud mass, $M_{\rm CO}$, calculated using $X_{\rm CO} = 3.0\times10^{20}$ cm$^{-2}$ (K km s$^{-1}$)$^{-1}$ versus minimal mass, $M_{min}$, required for tidal stability using the rotation curve from: (a) Feitzinger (1979), (b) Wong et al. (2008) and (c) Alves $\&$ Nelson (2000). The black 1-to-1 line is the locus of tidal balance.}   
  \label{fig:Annexe2}
\end{center}
\end{figure*}
\end{appendix}

\label{lastpage}
\end{document}